\documentclass[12pt]{article}

\usepackage{thumbpdf,lmodern}
\usepackage{epsfig}
\usepackage{graphicx}
\usepackage{amsmath, amsfonts, amsthm,amssymb}
\usepackage{latexsym}
\usepackage{color}
\usepackage{mathrsfs}
\usepackage{natbib}
\usepackage{hyperref}
\usepackage{appendix}
\usepackage{dsfont}
\usepackage[left=3cm,top=4cm,right=3cm,bottom=3cm]{geometry}
\RequirePackage[caption=false]{subfig}
\usepackage{lscape}
\usepackage{dsfont}
\usepackage{rotating}

\usepackage{mathabx}
\usepackage{calligra}

\usepackage{booktabs,tabularx,graphicx}
\usepackage{amsmath, amssymb}
\usepackage{dsfont}
\usepackage{float}
\usepackage{latexsym}
\usepackage{mathrsfs}
\usepackage{placeins}
\usepackage{wasysym}

\usepackage{framed}

\newcolumntype{C}{>{\centering\arraybackslash}X}

\newcommand{\plot}[1]{%
    \includegraphics[width=0.3\textwidth, height=0.3\textwidth, keepaspectratio]{#1}
}

\newcommand{\graph}{\mathcal{G}}
\newcommand{\node}{\mathcal{V}}
\newcommand{\edge}{\mathcal{E}}
\newcommand{\adj}{\mathbf{A}}
\newcommand{\var}{\mathds{V}ar}
\providecommand\longrightarrowrhd{\relbar\joinrel\relbar\joinrel\mathrel\rhd}

\begin{document}

\title{\LARGE \bf intensitynet: Intensity-based Analysis of Spatial Point Patterns Occurring on Complex Networks Structures
in \texttt{R}}
\date{\date{}}
\maketitle
\begin{center}
{{\bf Pol  Llagostera$^{a}$, Carles Comas$^{a}$} and {\bf Matthias Eckardt}$^{b}$}\\
\noindent $^{\text{a}}$ Department of Mathematics, Universitat de Lleida, Lleida, Spain\\
\noindent $^{\text{b}}$ Chair of Statistics, Humboldt-Universit{\"a}t zu Berlin, Germany\\
\end{center}
\begin{abstract}
The statistical analysis of structured spatial point process data where the event locations are determined by an underlying spatially embedded relational system has become a vivid field of research. Despite a growing literature on different extensions of point process characteristics to linear network domains, most software implementations remain restricted to either directed or undirected network structures and are of limited use for the analysis of rather complex real-world systems consisting of both undirected and directed parts. 
Formalizing the network through a graph theoretic perspective, this paper discusses a complementary approach for the analysis of network-based event data through generic network intensity functions and gives a general introduction to the \textbf{intensitynet} package implemented in \texttt{R} covering both computational details and applications. By treating the edges as fundamental entities, the implemented approach allows the computation of intensities and other related values related to different graph structures containing undirected, directed, or a combination of both edges as special cases. The package includes characteristics for network modeling, data manipulation, intensity estimation, computation of local and global autocorrelation statistics, visualization, and extensions to marked point process scenarios. All functionalities are accompanied by reproducible code examples using the \texttt{chicago} data as toy example to illustrate the application of the package.
\end{abstract}

\section{Introduction}

The statistical investigation and characterization of point configurations in event-type data on relational objects, i.e. spatially embedded network structures, have attracted a lot of attention in recent years. Typical examples include the locations of accidents or crimes in traffic systems, insects on brick structures, and spines in neural networks. In any such data, the event locations are only observed on or along the individual structural entities (i.e. the edges)  which, in turn, constitute the relational system of interest within a planar observations window. Dating back to the pioneering contributions on event-type data on linear networks of  \cite{https://doi.org/10.1111/j.1538-4632.1995.tb00341.x} and \cite{https://doi.org/10.1111/j.1538-4632.2001.tb00448.x}, \cite{10.2307/23357213} were the first to consider geometrically corrected summary characteristics for the analysis of network-based event patterns that account for the inherent structural properties of the relational system under study. Extending this idea further, a rich body of the literature has been established on network-adjusted extensions of classical spatial point process characteristics to spatially embedded structured domains   \citep[see][for a recent review of the literature]{BADDELEY2021100435}. While extensions to 
 directed linear networks \citep{Rasmussen2021} and graphs with Euclidean edges \citep{10.1214/19-AOS1896, LieshoutNets} were proposed, most tools are of limited use for the analysis of more complex, real-world network structures where both directed and undirected edges coexist. In particular, the \textbf{spatstat} \citep{spatstat:book, spatstat:JSS}, \textbf{DRHotNet} \citep{DRHotNet}, \textbf{geonet} \citep{geonet},  \textbf{spNetwork} \citep{spnet:R, spnet:git}, \textbf{SpNetPrep}\citep{SpNetPrep} and 
\textbf{stlnpp} \citep{stlnpp} packages in \texttt{R} \citep{RCore}
 do not help for the joint analysis of events on different types of edges nor the computation of point characteristics at different structural entities, e.g. set of neighbors or paths, of the network.   

Addressing these limitations, \cite{Eckardt2018, Eckardt2021} introduced a complementary approach that formalizes the network-based event data through different  \emph{network intensity functions} using a graph theoretic perspective. Treating the edges as fundamental entities, this approach allows for the computation of intensities and related quantities over different types and graph theoretic sets of (hyper-)structural entities including directed, undirected, or combinations of different, i.e. mixed, edges. All event-based characteristics derive directly from the intensities of the corresponding edges included where each edgewise intensity corresponds to the number of events, i.e. count per edge adjusted for its length. Although  \cite{Eckardt2018, Eckardt2021} derived a rich theoretical framework that allows for the analysis of event-type data on different network structures, no general implementation of the methodological toolbox exists. Summarizing their network intensity functions approach into an open-source solution in \texttt{R}, the \textbf{intensitynet} package is designed to fill this gap.

Implemented through \texttt{S3} classes in a  modular programming structure, the package allows for easy computation of edgewise, nodewise, and pathwise characteristics for event-type data on (sets of) directed, undirected, and mixed-type network structures. To ease its handling, the package is built around one constructor function, i.e.  \texttt{intensitynet()}, which automatically parses the underlying network structure to all computations.  
Using the capacities of the \textbf{igraph} \citep{igraph} package, all results are provided as edge or node attributes which allow for easy manipulation and retrieval of the stored information. Apart from summary tables, the package includes flexible visualization tools which help to produce (i) a reliable plot of the network structure itself, i.e. the observed patterns on the paths between any two nodes (origin and destination), and (ii) outputs from the computed intensity functions, the corresponding mark proportions or averages, and autocorrelation statistics. Each plot can be personalized with a few arguments allowing, among other things, to show only certain nodes or edges, deciding to show or hide the event locations, and selecting its transparency. 
The package is available from the Comprehensive R Archive Network (CRAN)
at \url{https://cran.r-project.org/package=intensitynet} and also as development version via the repository \url{https://github.com/LlagosteraPol/IntensityNet/}.

This paper provides a general introduction to the \textbf{intensitynet} package including data manipulation, intensity estimation, the computation of local and global autocorrelation statistics,  visualization, and extensions to marked point process scenarios. The functionalities of the package are tested and exemplified using the \texttt{chicago} dataset provided by the \textbf{spatstat} package.

\section{Structure and main functionalities}

Embedded into a graph theoretic framework, the \textbf{intensitynet} package translates the observed network structure into a graph $\graph=(\node, \edge)$ with $\node=\lbrace v_i \rbrace^n_{i=1}$ and $\edge=\lbrace e_j\rbrace^m_{j=1}\subseteq \node\times \node$ denoting the sets of vertices and edges, respectively. 
 Each edge in $\edge$ is associated with a pair of (not necessarily distinct) vertices, i.e. its endpoints, which constitute the relational structure of $\graph$. For a traffic network, the edges could correspond to the set of distinct road segments and the vertices to the intersections of the edges, e.g. the crossings in the traffic system. Formally, the observed network structure is translated into a graph object through a $n \times n$ matrix $\adj$ with element   $a_{i,j}$. Each element $a_{i,j}$ numerically reflects the presence or absence of an edge joining $v_i$ and $v_j$ through zero and nonzero values with zeros indicating missing edges. Usually, a binary coding scheme is applied such that $\adj$ only contains zeros and ones with ones corresponding to the edges contained in $\edge$. We note that apart from this binary specification, more challenging coding schemes exist for weighted graphs in which weight $\omega$, i.e. numerical attribute, is assigned to either the nodes or the edges, and multigraphs, where sets of vertices are joined by multiple edges. Under the present implementation of the \textbf{intensitynet} package, any such non-binary values need to be transformed into binary values in a preprocessing step. 
 
If the road segments are direction preserving, i.e. if movement along a particular road is only possible in one way, the corresponding edges are called \emph{directed}. In any such case, the associated vertices are defined by the ordered pair $(v_i,v_j)$ and the corresponding edges are indicated by an arc with head $v_j$ and tail $v_i$ leading to asymmetric $\adj$ as only $a_{i,j}$ but not $a_{j,i}$ is nonzero.
In contrast, any edge which does not impose any movement restrictions is called \emph{undirected} and related to the set $\lbrace v_i,v_j \rbrace$ corresponding to nonzero entries in both $a_{i,j}$ and $a_{j,i}$.  Different from the directed case, undirected edges are represented by lines. Depending on the edges, $\graph$ is called directed, if all edges are directed, undirected if all edges are undirected, and mixed if $\graph$ consists of both directed and undirected edges. See Table \ref{tab:net:examples} for a visual representation and corresponding specification of $\adj$ for all three possible graph representations covered by the \textbf{intensitynet} package.
\begin{table}[htb]
	\begin{center}
\setlength\tabcolsep{2pt}%
		\begin{tabularx}{\textwidth}{@{}c|*{3}{C}@{}}
		\hline
		\textbf{  Graph Type  } & \textbf{Graph Structure} & \textbf{   Adjacency Matrix} \\
		\hline
		Undirected &
		\vspace{4pt} \plot{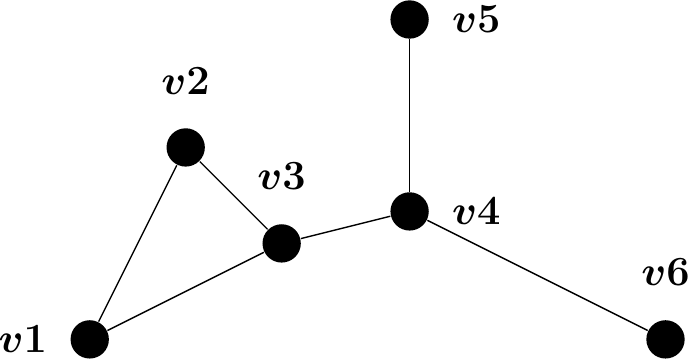} & 
		        \vspace{4pt}
				\(\adj=\begin{pmatrix}
                    0 & 1 & 1 & 0 & 0 & 0\\
                    1 & 0 & 1 & 0 & 0 & 0\\
                    1 & 1 & 0 & 1 & 0 & 0\\
                    0 & 0 & 1 & 0 & 1 & 1\\
                    0 & 0 & 0 & 1 & 0 & 0\\
                    0 & 0 & 0 & 1 & 0 & 0\\
                \end{pmatrix}\)
                \vspace{4pt}
                \\
        \hline
		\hline
		Directed &
		\vspace{4pt} \plot{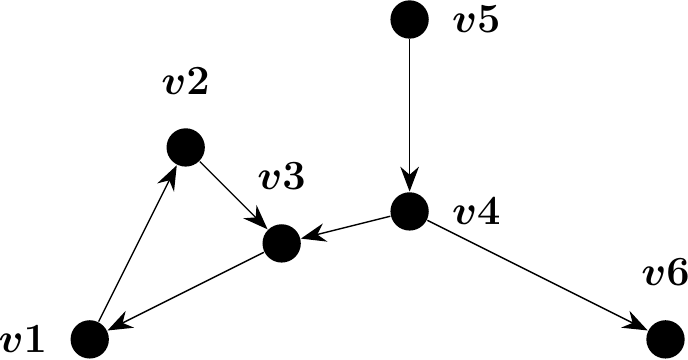} &  
		        \vspace{4pt}
		        \(\adj=\begin{pmatrix}
                    0 & 1 & 0 & 0 & 0 & 0\\
                    0 & 0 & 1 & 0 & 0 & 0\\
                    1 & 0 & 0 & 0 & 0 & 0\\
                    0 & 0 & 1 & 0 & 0 & 1\\
                    0 & 0 & 0 & 1 & 0 & 0\\
                    0 & 0 & 0 & 0 & 0 & 0\\
                \end{pmatrix}\) 
                \vspace{4pt}
                \\
        \hline
		\hline
		Mixed &
		  \vspace{4pt} \plot{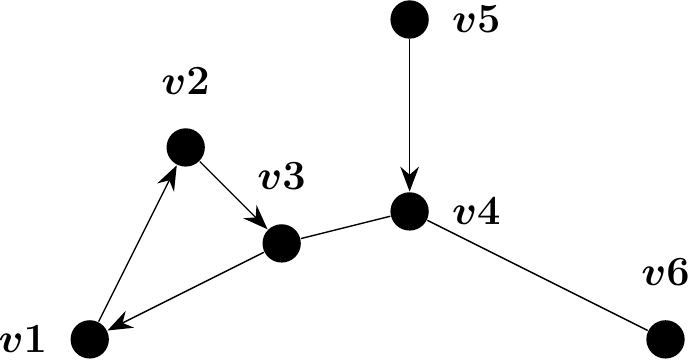} &
		        \vspace{4pt}
                \(\adj=\begin{pmatrix}
                    0 & 1 & 0 & 0 & 0 & 0\\
                    0 & 0 & 1 & 0 & 0 & 0\\
                    1 & 0 & 0 & 1 & 0 & 0\\
                    0 & 0 & 1 & 0 & 0 & 1\\
                    0 & 0 & 0 & 1 & 0 & 0\\
                    0 & 0 & 0 & 1 & 0 & 0\\
                \end{pmatrix}\) 
                \vspace{4pt}
                \\
        \hline
		\end{tabularx}
	\end{center}
	\caption{Examples of potential graph specifications and corresponding adjacency matrices covered by the \textbf{intensitynet} package}
	\label{tab:net:examples}
\end{table}
To relate the spatial network structure to its graph theoretical representation and investigate the distributional characteristics of the event locations over the edges, both the graph and the events need to be augmented by additional geographical information. To this end,  the sets of $n$ nodes and $p$ event locations are both reformulated into sets of coordinates $\lbrace\mathbf{v}_1,\ldots,\mathbf{v}_n)$ and $\lbrace \boldsymbol{\xi}_1,\ldots,\boldsymbol{\xi}_p\rbrace$ where $\mathbf{v}_i=(x_i,y_i)$ and  $\boldsymbol{\xi}_j=(u_i,w_j)$ are the exact coordinates of the $i$-th  node and $j$-th event with respect to a given coordinate reference system (CRS), respectively. As such, only the set of $n$ nodes $\lbrace\mathbf{v}_1,\ldots,\mathbf{v}_n)$ is treated as fixed while the $p$ events are considered as realizations of an unobserved stochastic mechanic which governs the locations on the network. Although both pairs of coordinates are treated differently,  both sets of coordinates must be derived from a unique CRS system. 

To explicitly control for the structured nature of the data and consider the points and network structure simultaneously, all summaries and characteristics in the \textbf{intensitynet} package derive directly from local computations, treating the edges as core elements. To this end, the edges are considered as edge intervals with associated Euclidean length which takes $(\mathbf{v}_i,\mathbf{v}_j)$ as endpoints. The edgewise intensity itself can then be computed as the sum over all events $\lbrace\boldsymbol{\xi}_p\rbrace$ that fall onto the interval spanned between $(\mathbf{v}_i,\mathbf{v}_j)$ adjusted for its length. 
All alternative nodewise and pathwise summaries included in the \textbf{intensitynet} package derive directly from the edgewise characteristics including directed, undirected and mixed graph versions
\citep[see][for detailed discussion]{Eckardt2018, Eckardt2021}. To correct for small spatial deviations of the locations of the events $\lbrace\boldsymbol{\xi}_p\rbrace$ from the edges, the \textbf{intensitynet} packages allow to specify a maximal spatial error distance between the edges and events, i.e. a spatial buffer, that assigns each event to a distinct edge based on the distance argument specified by the user (see Section \ref{sec:edgeeventasign} for detailed description).


\section{Operations on data}\label{sec:operations:data}

This section provides a detailed description of the different functionalities included in the  \textbf{intensitynet} package and explains its application using the \texttt{chicago} data,  originally provided by the \textbf{spatstat} \citep{spatstat:book, spatstat:JSS} package, as toy example.
To highlight its use in different graph settings, the \textbf{intensitynet} package provides three adapted versions of the original data, i.e. \texttt{und\_intnet\_chicago},  \texttt{dir\_intnet\_chicago}, and \texttt{mix\_intnet\_chicago}, based on undirected, directed and mixed graph specifications.

\subsection{Structure and classes}
\begin{figure}[h!]
	\begin{center}
		\includegraphics[scale=0.8]{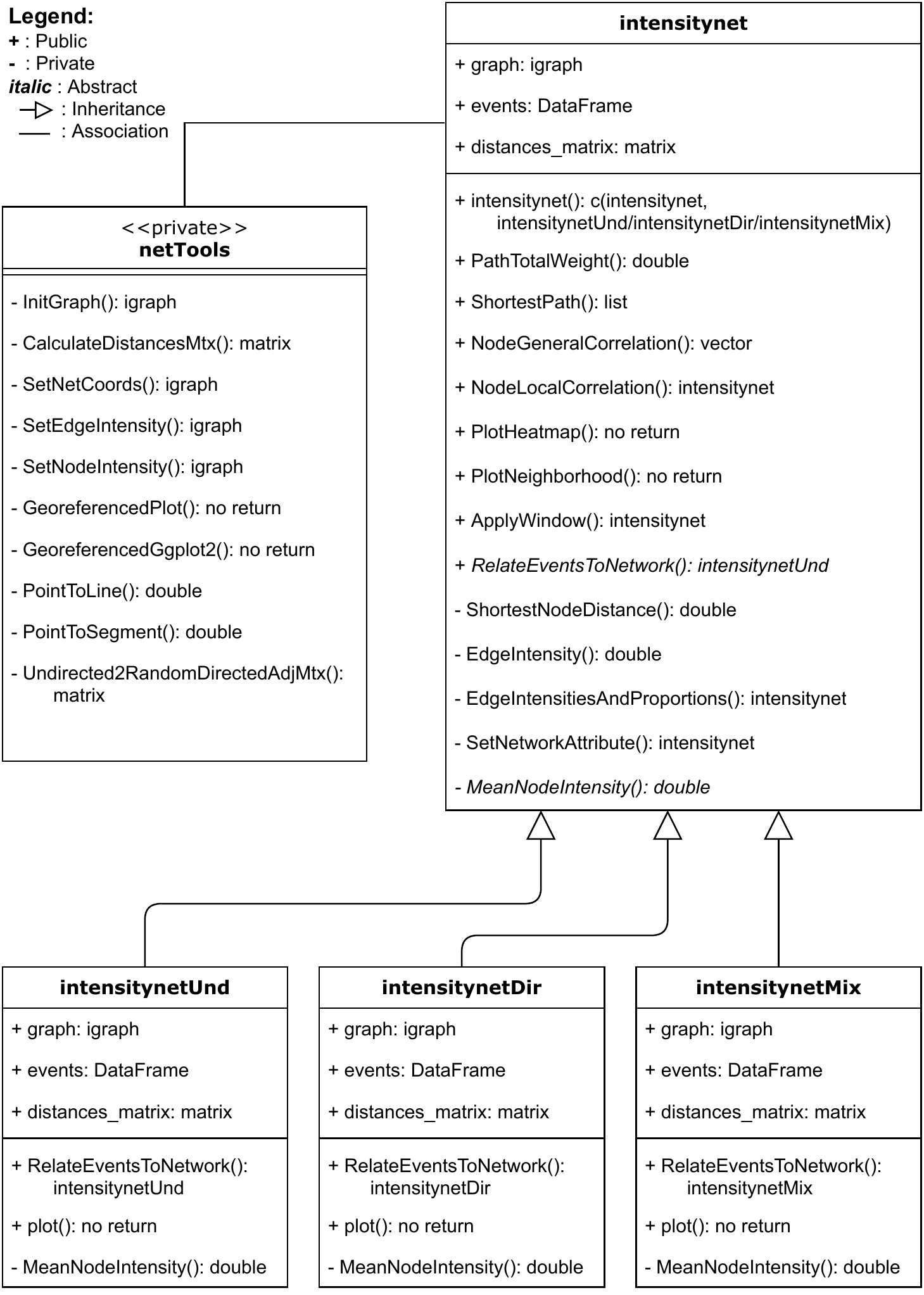}
	\end{center}
	\caption{UML class diagram of the inherent structure of the \textbf{intensitynet} package indicating its classes, methods, and internal relations. Functions provided by the different classes use the \textbf{-} and \textbf{+} symbols  to differentiate between private and public functions, respectively, and \textbf{\textit{italic}} to indicate abstract functions. Associations between the classes are indicated by solid lines and inheritance relations by arcs ($\longrightarrowrhd$).}
	\label{fig:uml:diagram}
\end{figure}

To introduce the basic operations and the interrelations among the functionalities of the \textbf{intensitynet} package, its internal structure needs to be described first. Conceptualized into two main and three subclasses, the  \textbf{intensitynet} package provides several functions (computational tools) to calculate, manipulate and visualize structured event-type data observed on relational systems. Its internal structure and the relation among the distinct (sub)classes and functions are presented in Figure \ref{fig:uml:diagram} in form of a Unified Modeling Language (UML) class diagram. At its core, the package creates a proper \texttt{intensitynet} object based on the given arguments of the \texttt{intensitynet()} function, and a  constructor associated with the main class \texttt{intensitynet} of the package. All general functionalities of the main class are provided as visible (public) functions. Specific methods for undirected, directed and mixed graphs are given by the three subclasses \texttt{intensitynetUnd}, \texttt{intensitynetDir}, and \texttt{intensitynetMix}  which inherit all methods from the main class. Associated with the \texttt{intensitynet} class, the package consists of a second main class, called \texttt{netTools}, which is constructed as a helper class with a set of functions used by the \texttt{intensitynet} class and corresponding graph specific subclasses. All methods of the helper class are implemented as invisible, i.e. private, functions and primarily designed for internal operations and should not be applied directly by the user. We note that apart from the main functionalities, both the \texttt{intensitynet} class and its related subclasses also contain private functions which are not designed for direct application.
To model, manipulate and store the network and to externally visualize the results, all network-based information and computational results including distances, intensities, and correlations are stored as node and edge attributes using the capacities of the \textbf{igraph} \citep{igraph}  package.  

\subsection{The intensitynet function}\label{sec:data:prep}

To apply any operations on network-based event data, the event locations and corresponding information derived from the graph theoretic reformulation of the underlying relational system needs to the transformed into an \texttt{intensitynet} object. This object is initialized by the constructor \texttt{intensitynet()} which, in turn, requires the following three elements as mandatory attributes:
\begin{itemize}
	\item \texttt{adjacency\_mtx}: a binary adjacency matrix of dimension $n\times n$  which specifies the network structure as the relation (adjacency) between each pair of nodes according to Table \ref{tab:net:examples}. 
	
	\item \texttt{node\_coords}: a $n \times 2$ vector as 
 \texttt{DataFrame} corresponding to the $n$ pairs of exact coordinates provided by $\lbrace \mathbf{v}_i\rbrace^n_{i=1}$. These coordinates will be used to calculate the distances of edges and to draw the network in the desired spatial structure (shape).
	
	\item \texttt{event\_data}: a $p \times 2$ vector as 
 \texttt{DataFrame} corresponding to the $p$ pairs of exact coordinates provided by $\lbrace \boldsymbol{\xi}_j\rbrace^p_{j=1}$. If additional marks, i.e. qualitative or quantitative point attributes,  are available, \texttt{event\_data} is required to be a \texttt{DataFrame} of dimension $p \times 3$ with the exact coordinates placed in the first two, and the mark information in the third columns.  The mark column could contain mixtures of real-valued (numerical) and categorical (string) entries. However, at each row, only one numerical or categorical mark is permitted.
\end{itemize}

For both \texttt{node\_coords} and  \texttt{event\_data}, the coordinates are required to be numbers without any punctuation or spaces. \\

In addition, two optional arguments can be provided to the \texttt{intensitynet()} function to customize the network structure and event-on-edge alignment. By default, \texttt{intensitynet()} assumes an undirected graph. For alternative graph structures, specification of the \texttt{graph\_type} is required to compute the corresponding set of suitable network intensity functions.

\begin{itemize}
    \item \texttt{graph\_type}: A string (text) indicating the specific network structure under study. Potential statements include \texttt {undirected}, treated as the default value, \texttt {directed}, and \texttt {mixed}. 
	
	\item \texttt{event\_correction}: Numerical value specifying the maximal distance in meters used to compute the event-to-edge alignment. By default, a maximal pointwise distance of $5$ meters is applied using the world cartography representation WGS84 as the coordinate reference system. It is strongly advised to carefully adapt this value to the CRS of the node and event locations. Higher values of the \texttt{event\_correction} argument potentially yield  an increase of (mis-)assigned events while values close to $0$ may increase the number of events that are not assigned to any edge in the network and, whence, excluded from any subsequent computations.
\end{itemize}

The constructor function returns a two-class object, i.e. an object with class \texttt{intensitynet} and corresponding subclass depending on the given \texttt{graph\_type} argument according to Figure \ref{fig:uml:diagram}. The secondary classes are used internally to call the corresponding functions and operations for undirected, directed, or mixed network structures. The resulting object wraps $4$ values into a list with the following information:

\begin{itemize}
	\item \texttt{graph}: the graph representation of the observed network stored as \texttt{igraph} object with the Euclidean length placed as an edge attribute called  \emph{weight}.
	
	\item \texttt{events}: DataFrame of events as provided by the user in the \texttt{event\_data} argument of \texttt{intensitynet} function.
	
	\item \texttt{graph\_type}:  A string corresponding to the specification of the graph type (\texttt{undirected}, \texttt{directed} or \texttt{mixed}).
	
	\item \texttt{distances\_matrix}: A $n\times n$ dimensional matrix with the pairwise Euclidean distances between the node locations on the network specified as the length of the virtual line joining any pair of nodes in the network.
	
	\item \texttt{event\_correction}: Value specified in the \texttt{event\_correction} argument of the \texttt{inten\-sitynet} function corresponding to the maximal distance used in the computation for the pointwise event-to-edge assignment.
\end{itemize}

All elements of the \texttt{intensitynet} object can directly be addressed, analogous to standard data operations in \texttt{R}, by using e.g. the  `$\$$´  symbol. 

\subsection{Specifying the intensitynet objects from alternative network formats}

Before providing a detailed treatment of available data manipulation operations in the \textbf{intensitynet} package, we briefly outline potential steps to compute the \texttt{intensitynet} function from (i) a linear point process (\texttt{lpp}) object as provided by the \textbf{spatstat} package and (ii) shapefile (\texttt{shp}) information.  Although the aim of our package is not to read, adapt or convert different formats into an \texttt{intensitynet} object, all mandatory arguments of the \texttt{intensitynet()} function can be retrieved from different sources using a variety of available \texttt{R} packages. 

\subsubsection{Specifying the intensitynet object from  linear point process object}

In what follows, the main steps to gather all relevant information, i.e. the coordinates of the nodes, the corresponding adjacency matrix $\adj$, and the set of event locations, from a \texttt{lpp} object, using the capacities of the \textbf{igraph} package are described and illustrated using the \texttt{chicago}  data of the \textbf{spatstat} package.\\ 

In the \textbf{spatstat} package, a \texttt{lpp} object is jointly defined from two objects: (i) a point pattern object (\texttt{ppp}) which specifies the observed event locations and (ii) a corresponding linear network object (\texttt{linnet}) which specifies the underlying network structure as a network of line segments. Having installed and loaded the \textbf{spatstat} and \textbf{intensitynet} packages and   \texttt{chicago}  data into the \texttt{R} environment, information on the edges can be extracted from the \texttt{from} and \texttt{to} vectors of the \texttt{domain} attribute of the \texttt{linnet} object which specifies the individual edges by the corresponding pairs of endpoints (\texttt{from}, \texttt{to}). The exact coordinate information for the vertices can then be retrieved from the \texttt{x} and \texttt{y} vectors of the \texttt{vertices} attribute of the \texttt{domain} object. Executing the following lines, both the edges and the vertices are stored as a \texttt{vector} and \texttt{DataFrame}, respectively.

\begin{verbatim}
R> library(intensitynet)
R> library(spatstat)
R> data(chicago)

R> edges <- cbind(chicago[["domain"]][["from"]], chicago[["domain"]][["to"]])
R> node_coords <- data.frame(xcoord=chicago[["domain"]][["vertices"]][["x"]], 
+                            ycoord=chicago[["domain"]][["vertices"]][["y"]])
\end{verbatim}

Given the coordinates of the nodes, the second mandatory argument of the \texttt{intensitynet()} function, i.e. the adjacency matrix $\adj$, can be computed directly from the \texttt{edges} object using subsequently the \texttt{graph\_from\_edgelist()} and \texttt{as\_adjacency\_matrix()} functions from the  \textbf{igraph} package. 

\begin{verbatim}
R> net <- igraph::graph_from_edgelist(edges)
R> adj_mtx <- as.matrix(igraph::as_adjacency_matrix(net))
\end{verbatim}

Finally, ignoring any additional mark information provided by the  \texttt{chicago} data, the event locations can be retrieved from the \texttt{x} and \texttt{y} columns of the \texttt{data} attribute of the \texttt{lpp} object. We note that identical coordinates information to \texttt{x} and \texttt{y} is also provided in the columns \texttt{seg} and \texttt{tp}, but using an alternative coordinate reference system. Collecting the required information, the \texttt{intensitynet} object can then be specified by executing the following lines. 

\begin{verbatim}
R> chicago_df <- as.data.frame(chicago[["data"]][, -(3:4)]) 

R> intnet_chicago <- intensitynet(adjacency_mtx = adj_mtx, 
+                                 node_coords = node_coords, 
+                                 event_data = chicago_df)	
R> attributes(intnet_chicago)
$names
[1] "graph"   "events"   "graph_type"   "distances_mtx"   "event_correction"

$class
[1] "intensitynet"    "intensitynetUnd"
\end{verbatim}

Compared to the required operations of the \textbf{spatstat} package for network event data importation, the \textbf{intensitynet} appears to be highly flexible with eager object generability. Different from the simple network object specification through the \texttt{intensitynet()} function and essential data requirement, \textbf{spatstat} asks to perform several data transformations to define a linear network pattern. In detail, this process requires to subsequently create a point process object (\texttt{ppp}) and a \texttt{linnet} object which itself derives from an \texttt{ppp} object plus a two-column matrix specifying the edges and, finally, the linear point pattern object (\texttt{lpp}) with the \texttt{linnet} object plus the location of the points (coordinates). Moreover, if it is intended to create an \texttt{igraph} object using the linear point pattern, more conversions are required using the help of additional libraries such as \texttt{sf} and \texttt{sfnetworks}.


\subsubsection{Specifying the intensitynet object from  shapefile data}\label{sec:shapefile}

Shapefile (\texttt{shp}) event-type data emerge in different contexts, in particular in public data, and is commonly used in cartographic tools such as ArcGIS \citep{ArcGis:software} or QGIS \citep{QGIS:software}. As for the \texttt{lpp} object, any \texttt{shp} object can be transformed into the desired formats using the capacities of the \textbf{maptools} \citep{maptools} and \textbf{shp2graph} \citep{Shp2graph} packages. All required steps are illustrated using the \texttt{ORN.nt} data on the Ontario road network \citep{ontario:2006} provided by the \textbf{shp2graph}  package.\\
\\
If required, any shapefile could  initially be read  into the \texttt{R} environment using the function \texttt{readShapeLines()} from \textbf{maptools} which stores the provided information as \texttt{SpatialLines\-DataFrame} object, an object of class \texttt{sp}. Here, as the \texttt{ORN.nt} data is already provided in \texttt{sp} format, this initial step through the \textbf{maptools} package  is omitted. Given a \texttt{sp} object as in our example, the object can be transformed into a list using the \texttt{readshpnw()} function of the \textbf{shp2graph}. By setting the argument \texttt{ELComputed} of the \texttt{readshpnw()} function to \texttt{TRUE},  the list will also contain the edge lengths.

\begin{verbatim}
R> library(intensitynet)
R> library(shp2graph)
R> data(ORN)

R> rtNEL <- shp2graph::readshpnw(ORN.nt, ELComputed = TRUE) 
\end{verbatim}

Executing the above code yields a list with the following elements: a Boolean stating if the list is in  \emph{Detailed}  mode (\texttt{FALSE} by default), the list of nodes from the network, the list of edges, the lengths of the edges, a DataFrame of edge attributes extracted from the \texttt{sp} object, a vector with the \texttt{x} coordinates of all the nodes, and a vector containing the \texttt{y} coordinates. From this list, only information on the nodes, edges, and lengths of the edges are needed for further operations. Using the \texttt{nel2igraph()} from the  \textbf{shp2graph} package, this selected information can finally be translated into an \texttt{igraph} object as outlined in the following lines.

\begin{verbatim}
R> nodes_orn <- rtNEL[[2]] 
R> edges_orn <- rtNEL[[3]] 
R> lenghts_orn <- rtNEL[[4]]

R> net_orn<-shp2graph::nel2igraph(nodelist = nodes_orn,
+                                 edgelist = edges_orn,
+                                 weight = lenghts_orn)
\end{verbatim}

From the \texttt{net\_orn} object, all required information to initialize the \texttt{intensitynet()} function can easily be obtained. Noting that \texttt{net\_orn}  object is a network in \texttt{igraph} format, its adjacency matrix $\adj$ can directly be recovered using the \texttt{graph\_from\_edgelist()} and \texttt{as\_adjacency\_matrix()} functions as outlined in the previous section.  The coordinates of the nodes can be extracted from the \texttt{nodes\_orn} object using the \texttt{Nodes.coordinates()} function from the  \textbf{shp2graph} package. Given the above information, the initialization of the \texttt{intensitynet} object is exemplified below. Notice that the argument \texttt{event\_data} is an empty matrix with two columns (corresponding to the number of columns required) as no events are available in the original data.

\begin{verbatim}
R> adj_mtx_orn <- as.matrix(igraph::as_adjacency_matrix(net_orn))
R> node_coords_orn <- shp2graph::Nodes.coordinates(nodes_orn)

R> intnet_orn <- intensitynet(adjacency_mtx = adj_mtx_orn, 
+                             node_coords = node_coords_orn, 
+                             event_data = matrix(ncol = 2))
\end{verbatim}

\subsection{Event-to-edge alignment and network intensity function estimation}\label{sec:data:manipulation}

\subsubsection{Event-to-edge alignment}\label{sec:edgeeventasign}

The \texttt{intensitynet} object described in Section \ref{sec:data:prep} summarizes the network-based event data together with information on the network as such including the node coordinates, its associated adjacency matrix, and the edge distances. Although the \texttt{intensitynet} object already allows for the direct computation of different graphical outputs and to retrieve the data or \texttt{igraph} object, it does not provide any event characteristics, i.e. intensity functions or autocorrelation statistics. To derive any further information on the event pattern from the \texttt{intensitynet} object, the individual event locations $\lbrace\boldsymbol{\xi}_j\rbrace_{j=1}^p$ need to be assigned to the edges in a subsequent action using the \texttt{RelateEventsToNetwork()} function. At its core, this function computes the distances $d_{j,k}$ of the $j$-th event to all $k$ edges in the graph and assigns the event to exactly one edge based on the minimal distance $d_{j,k}$ which, in turn, depends on a pre-specified threshold distance $\tau$ stated by the user in the \texttt{event\_correction} argument of the  \texttt{intensitynet} object. 

\begin{figure}[htb]
	\begin{center}
		\begin{tabular}{ccc}
			\scalebox{0.96}{\includegraphics[scale=0.75]{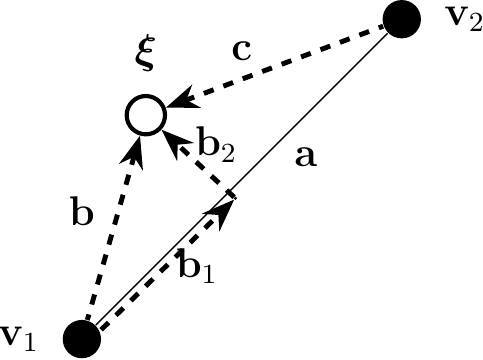}} & 
			\scalebox{1} {\includegraphics[scale=0.75]{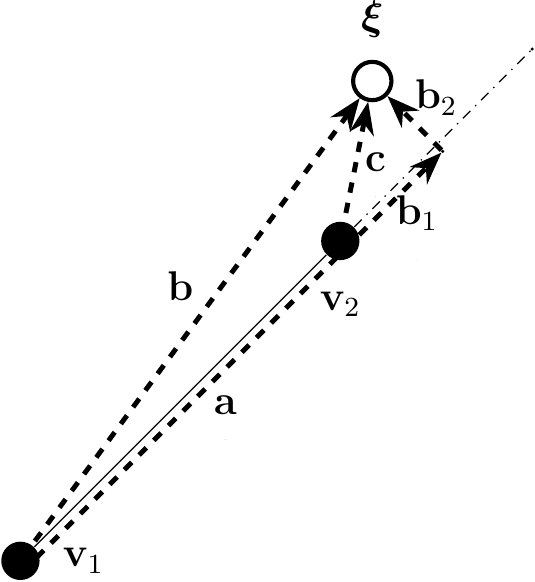}}& 
			\scalebox{1}{\includegraphics[scale=0.75]{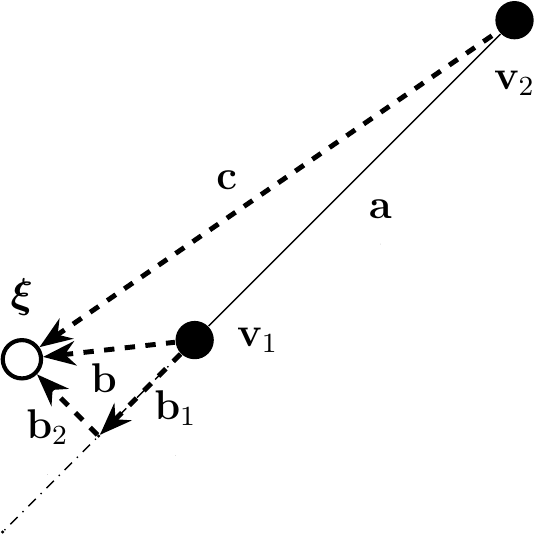}} \\
			$\graph_1$ & $\graph_2$ & $\graph_3$ \\
		\end{tabular}
	\end{center}
	\caption{Three potential scenarios encountered in the event-to-edge alignment. The event $\boldsymbol{\xi}$ is located between the nodes $\mathbf{v}_1$ and $\mathbf{v}_2$ of $\graph_1$ (left), the  event $\boldsymbol{\xi}$  is not located between $\mathbf{v}_1$ and $\mathbf{v}_2$ but close to $\mathbf{v}_2$ of $\graph_2$ (central), and the event $\boldsymbol{\xi}$  is not located between $\mathbf{v}_1$ and $\mathbf{v}_2$ but close to $\mathbf{v}_1$ of $\graph_3$ (right).}
	\label{fig:projection:example}
\end{figure}

To illustrate the potential event-to-edge assignment scenarios, consider the three toy graphs $\graph_1, \graph_2$ and $\graph_3$ with nodes $\mathbf{v}_1$ and  $\mathbf{v}_2$ denoting the endpoints of an edge and $\boldsymbol{\xi}$ an artificial event location that calls for an event-to-edge assignment. Mathematically, any such assignment operation can be defined through the distance $\delta(\boldsymbol{\xi}, (\mathbf{v}_1,\mathbf{v}_2))$  between the event $\boldsymbol{\xi}$ and its projection to an edge with endpoints $(\mathbf{v}_1, \mathbf{v}_2)$. Using the distance $\delta$ as decision criteria, any event $\boldsymbol{\xi}$ is re-assigned to its closest edge. Note that each event-to-edge assignment only accounts for the subset of edges satisfying $\delta\leq \tau$. Theoretically, three potential cases might appear as illustrated in the toy examples $\graph_1$ to $\graph_3$ (compare Figure \ref{fig:projection:example}). While the event $\boldsymbol{\xi}$ falls into a rectangular area formed by both nodes $\mathbf{v}_1$ and $\mathbf{v}_2$ in scenario one ($\graph_1$),  $\boldsymbol{\xi}$ is in the exterior but close to the area 
spanned by $\mathbf{v}_1$ or $\mathbf{v}_2$ in the two alternative scenarios. To decide on the closest distance $\delta$ to the edge $(\mathbf{v}_1,\mathbf{v}_2)$, the \textbf{intensitynet} package performs a triangulation based on a vector projection and rejection approach   \citep[see][for detailed discussion]{engineermaths} to select the optimal event-to-edge assignment. 

Formally, given the nodes $\mathbf{v}_1 = (x_1, y_1)$, $\mathbf{v}_2 = (x_2,y_2)$ and the event $\boldsymbol{\xi} = (u,w)$ on a network, simple calculation yield the vectors $\mathbf{a} = \mathbf{v}_2 - \mathbf{v}_1 = (x_2 - x_1, y_2 - y_1),~\mathbf{b} = \boldsymbol{\xi} - \mathbf{v}_1   = (u - x_1, w - y_1)$ and $\mathbf{c}=\boldsymbol{\xi} - \mathbf{v}_2   = (u - x_2, w - y_2)$. The definition of $\mathbf{a}$ and $\mathbf{b}$ allows to compute projection  $\mathbf{b}_1$  of  $\mathbf{b}$ onto the edge  $(\mathbf{v}_1,\mathbf{v}_2)$ where $\mathbf{b}_1 = ((\mathbf{b}\cdot \mathbf{a})/(\mathbf{a}\cdot \mathbf{a}))\times \mathbf{a}$ and $\cdot$ denotes the dot product of the vectors $\mathbf{a}$ and $\mathbf{b}$. Finally, from $\mathbf{b}$ and its projection $\mathbf{b}_1$, the vector rejection $\mathbf{b}_2$ can be obtained from calculation of 
$\mathbf{b}_2 = \mathbf{b} - \mathbf{b}_1$
 as $\mathbf{b}=\mathbf{b}_1+\mathbf{b}_2$.
 
Under the above formulation and depending on the exact location of $\boldsymbol{\xi}$ relative to $\mathbf{a}$, the closest edge could be either $\mathbf{b}$,  $\mathbf{b}_2$ or $\mathbf{c}$. To decide on which one is the closest, the vector distances of $\mathbf{b}_1$ and $\mathbf{a}$ can be used to determine whether $\boldsymbol{\xi}$ falls into the area spanned by  $\mathbf{v}_1$ and $\mathbf{v}_2$ or an exterior region close to $\mathbf{v}_1$ or $\mathbf{v}_2$. If the location of $\boldsymbol{\xi}$ is between $\mathbf{v}_1$ and $\mathbf{v}_2$, the closest vector is $\mathbf{b}_2$, if it falls in an exterior area but close to $\mathbf{v}_2$, the closest vector is $\mathbf{c}$, and if it falls in an exterior area but is close to $\mathbf{v}_1$, the closest vector is  $\mathbf{b}$. 

The following decision criteria on the event-to-edge assignment can be derived - assuming either positive (left) or negative (right) directions of $\mathbf{a}$:
\begin{align*}
0 < |\mathbf{b}_1| < |\mathbf{a}| \rightarrow \mathbf{b}_2&\text{~and~} 0 > |\mathbf{b}_1| > |\mathbf{a}| \rightarrow \mathbf{b}_2\\
0 \geq |\mathbf{b}_1| < |\mathbf{a}| \rightarrow \mathbf{b}&\text{~and~}0 \leq |\mathbf{b}_1| > |\mathbf{a}| \rightarrow \mathbf{b}\\
0 < |\mathbf{b}_1| \geq |\mathbf{a}| \rightarrow \mathbf{c}&\text{~and~}0 > |\mathbf{b}_1| \leq |\mathbf{a}| \rightarrow \mathbf{c}.
\end{align*}

Based on the above decision criteria, the distance to the closest selected edge is calculated. This is the distance that will be used to determine whether the event is accounted into the edge events. The described procedure is implemented in the private function \texttt{PointToSegment()} of the class \texttt{netTools} and used by the \texttt{RelateEventsToNetwork()} function .\\

\subsubsection{Network intensity function estimation}

Having performed an event-to-edge assignment for all events of the \texttt{intensitynet} object, the  \texttt{RelateEventsToNetwork()} function internally computes the edgewise and nodewise intensities of the network-based event pattern. As the core element of all network intensity function computations,  the edgewise intensity which quantifies the number of events for each edge in the network adjusted for the length of the edge is computed in an initial step. We note that different from all alternative statistics, the edgewise intensity function is calculated for all edges regardless of the type and potential direction restrictions. This local quantity is then used in subsequent actions to compute the nodewise and also pathwise intensity functions. The nodewise intensity is computed as the average of the edgewise ones over the set of edges included in the hyperstructural set under study, i.e. the set of neighbors, the set of parents, or the set of children   \citep[see][for the formulation and mathematical details on different network intensity functions]{Eckardt2018, Eckardt2021}.  

Internally, the edgewise and nodewise intensities are implemented in the private functions \texttt{EdgeIntensity()} and \texttt{MeanNodeIntensity()}, respectively. The \texttt{MeanNode\-Intensity()} function is an abstract method from the \texttt{intensitynet} class and is implemented by its subclasses \texttt{intensitynetUnd}, \texttt{intensitynetDir}, and \texttt{intensitynetMix} (see Figure \ref{fig:uml:diagram}). The second class of \texttt{intensitynet} object allows to automatically compute the correct nodewise intensity depending on the network type (undirected, directed, or mixed). For undirected networks, the \texttt{MeanNodeIntensity()} function yields only one output which is stored under the name  \texttt{intensity}. Different from the undirected case, application of the \texttt{MeanNodeIntensity()} function to directed graphs yields two different network intensity functions corresponding to the set of parent $pa$ (where $pa(i)= {j \rightarrow i} $) and set of children $ch$ (where $ch(i)={i \rightarrow j}$) that are stored under the names  \texttt{intensity\_in}  and  \texttt{intensity\_out},   respectively. Formally, the \texttt{intensity\_in} object is computed by averaging the edgewise intensities over the set of incident edges with respect to a given vertex. Likewise, the \texttt{intensity\_out} object reports the average intensity at node level constructed from the edgewise intensities of the set of dissident edges. Lastly, all these three outputs (\texttt{intensity\_und}, \texttt{intensity\_in},  \texttt{intensity\_out}) and one additional object called \texttt{intensity\_all} are provided in the case of mixed graphs, i.e. when the graph consists of both undirected and directed edges \citep[see][for a detailed review of Graph Theory concepts]{graphtheory}. The application of the \texttt{RelateEventsToNetwork()} function to empirical data is illustrated in the following lines. 

Continuing with the \texttt{chicago} toy example,  the internal event-to-edge assignment based on the \texttt{edge\_correction} arguments and subsequent intensity estimations are initialized by directly applying the \texttt{RelateEventsToNetwork()} function to the \texttt{intensitynet} object. Recalling that all computations are internally stored as edge and node attributes, the graph object \texttt{g} needs to be addressed to extract the stored information from the results. Given the graph object, both the nodewise and edgewise  intensities could be retrieved by specifying the \texttt{what} argument of the \texttt{igraph::as\_data\_frame(g, what)} command as \texttt{vertices} and \texttt{edges}, respectively.

\begin{verbatim}
R> intnet_chicago <- RelateEventsToNetwork(intnet_chicago)
Calculating edge intensities with event error distance of 5...
  |=============================================================| 100%
Calculating node intensities...
  |=============================================================| 100%

R> g <- intnet_chicago$graph
R> class(g)
[1] "igraph"

R> head(igraph::as_data_frame(g, what = "vertices"))
   name      xcoord   ycoord   intensity
V1   V1   0.3894739 1253.803 0.000000000
V2   V2 109.6830137 1251.771 0.000000000
V3   V3 111.1897363 1276.560 0.000000000
V4   V4 198.1486340 1276.560 0.000000000
V5   V5 197.9987626 1251.153 0.008242282
V6   V6 290.4787354 1276.560 0.000000000

R> head(igraph::as_data_frame(g, what = "edges"))[1:5]
  from  to    weight n_events  intensity 
1   V1  V2 109.31241        0 0.00000000   
2   V2  V3  24.83439        0 0.00000000  
3   V2  V5  88.31791        0 0.00000000   
4   V2 V24  54.86415        0 0.00000000   
5   V4  V5  25.40732        0 0.00000000   
6   V5  V7  90.99421        3 0.03296913   
\end{verbatim}

If the \texttt{intensitynet} object contains additional mark information, the \texttt{RelateEvents\-ToNetwork()} function automatically calculates additional edgewise summaries in addition to  the intensity computations. Accounting for the mark information provided, the  \texttt{RelateEventsTo\-Network()} function computes either edgewise averages (numerical marks) or proportions (categorical marks) of all events that fall into the edge interval considered. All computations including the edgewise and  nodewise intensities, the mark means and proportions, and the number of events per edge are stored as node and edge attributes using the \texttt{igraph} class.

To give an example of the underlying computations, consider a set of network-based events with numerical mark \emph{km} and categorical mark \emph{animals} with levels \emph{cat},  \emph{dog}, and \emph{cow}. For \emph{km} and \emph{animals}, suppose that we observed three events with marks $100, 20, 30$ and one \emph{cat} recorded twice and \emph{dog} appearing one time. Given the above specified data, application of the \texttt{RelateEventsToNetwork()} function results in the values $50$ for \emph{km}, $0.666$ for \emph{cat} and $0.333$ for \emph{dog}. As no event is labeled as \emph{cow} in the toy data, the output will include a zero in the corresponding field.

Given any empirical network-based event data with additional mark information, the internal edge-to-event assignment and subsequent calculation can be implemented analogously to the unmarked case as illustrated above. From the output of the \texttt{Relate\-EventsToNetwork()} function,  all results can be retrieved in subsequent actions directly from the graph object \texttt{g} as illustrated in the next lines.

\begin{verbatim}
R> head(igraph::as_data_frame(g, what = "edges"))[6:12]
  assault burglary cartheft    damage   robbery theft trespass
1       0        0        0 0.0000000 0.0000000     0        0
2       0        0        0 0.0000000 0.0000000     0        0
3       0        0        0 0.0000000 0.0000000     0        0
4       0        0        0 0.0000000 0.0000000     0        0
5       0        0        0 0.0000000 0.0000000     0        0
6       0        0        0 0.3333333 0.6666667     0        0
\end{verbatim}

Apart from working on the complete network, the \textbf{intensitynet} package allows the selection of smaller areas from the network using the  \texttt{ApplyWindow()} function. As before, this function only requires an \texttt{intensitynet} object as input and the pairs of $(x,y)$ coordinates specifying the boundaries of the selection window. The function extracts the induced sub-network along with corresponding event locations and returns an object with the same class type as its origin. We note that apart from the pre-selection of the observation window, the function also allows for post-selection of sub-networks. In that case, all computations corresponding to the selected area will be inherited from the sub-network. 

To illustrate the \texttt{ApplyWindow()} function, we outline its application for the \texttt{chicago} toy data example. Using the \texttt{intnet\_chicago} object as input, an extract of the substructure with borders $x=(300, 900)$ and $y=(500, 1000)$ is computed executing the next lines of code.  

\begin{verbatim}
R> sub_intnet_chicago <- ApplyWindow(intnet_chicago, 
                                     x_coords = c(300, 900), 
                                     y_coords = c(500, 1000)) 
\end{verbatim}

Given any substructure, both the original and the selected network event patterns can be compared using e.g. the \texttt{gorder()} and \texttt{gsize()} functions which compute the number of nodes and edges in the corresponding networks, respectively, as outlined next. 

\begin{verbatim}
R> c(igraph::gorder(intnet_chicago$graph),
      igraph::gsize(intnet_chicago$graph),
      nrow(intnet_chicago$events))
[1] 338 503 116   

R> c(igraph::gorder(sub_intnet_chicago$graph),
      igraph::gsize(sub_intnet_chicago$graph),
      nrow(sub_intnet_chicago$events))
[1] 75 112 37
\end{verbatim}

\subsection{Exploring the network structure}

Before proceeding with the estimation of alternative event-related characteristics, summaries of the network structure itself provided by \textbf{intensitynet} package are presented first.

Given any network, an important graph theoretic concept that helps to quantify the structural pairwise interrelations between the distinct  nodes is the shortest path \citep{Rahman2017}. From a graph theoretic perspective, a \emph{path} connecting any two vertices (origin and destination) is defined as a sequence of distinct vertices $(v_1, v_2, \cdots , v_n)$ with a maximum degree of $2$ and distinct edges $(e_1, e_2, \cdots e_{n-1})$ such that no edge and no vertex is traversed twice. A \emph{shortest path} between an origin and a destination is the minimum number of edges (if unweighted) or the minimum total weight of the edges (if weighted) to move along a path from the origin to the destination. Several algorithms and methodologies can be found in the literature that computes the shortest path including the algorithms of Dijkstra, Bellman-Ford, Floyd-Warshall, \citep{Magzhan:etal:2013}, and Johnson \citep{Johnson:etal:1977} or the Bread-First search algorithm to name just a few. Each of these algorithms serves better for certain networks with particular  characteristics and, in turn,  the choice of the algorithm used has a crucial impact on the computation and might yield an increase in the computational time or even intractable computations.
\\
The function \texttt{ShortestPath()} of the \textbf{intensitynet} package is designed to simplify the shortest path detection on the network. Basically, this function calculates the shortest path between two nodes based on either the number of edges or their weights. To operate, the function only requires the network of class \texttt{intensitynet}, the origin and destination nodes (either its names as strings or IDs as integers) specifying the path under selection, and an optional parameter specifying the weight type as arguments. If no weight is provided, the function calculates the shortest path based on the number of edges. If weights are available, the shortest path detection returns the minimum total weight (sum of all path edge weights). Regarding the chosen algorithm that calculates the shortest path, the function will choose automatically the one that fits best for the given network. Internally, the function performs a  selection based on the network characteristics. If the network is unweighted (or the weights are not considered) then the Bread-First search algorithm is used. If the user specifies a valid weight (that is present in the edge attributes), then Dijkstra's algorithm is chosen. For negative weights,   the shortest path detection either uses  Johnson's algorithm, if the network consists of more than $100$ nodes,  or Bellman-Ford's algorithm if the number of nodes is less than $100$.

The use of the \texttt{ShortestPath()} function and the specification of its argument are demonstrated in the next lines of code using the \texttt{intnet\_chicago} object. Selecting the nodes  $\mathbf{v}_1$ (\texttt{V1})  and  $\mathbf{v}_{300}$ (\texttt{V300}) from the  \texttt{intnet\_chicago} object, the following lines outline the shortest path detection with respect to the minimum number of edges, the minimum pathwise intensity, and the minimum number of car thefts taking additionally the mark information into account.

\begin{verbatim}
R> short_path <- ShortestPath(intnet_chicago, 
                              node_id1 =  "V1" , 
                              node_id2 =  "V300")
R> short_path$path
+ 16/338 vertices, named, from d8a2c9e:
 [1]  V1   V2   V24  V46  V65  V83  V96  V115 V123 V125 V134 V219 V220 
 [14] V221 V294 V300

R> short_path$total_weight
[1] 16

R> short_path_intensity <- ShortestPath(intnet_chicago, 
                                        node_id1 =  "V1" , 
                                        node_id2 =  "V300" , 
                                        weight =  "intensity")
R> short_path_intensity$path
+ 44/338 vertices, named, from d8a2c9e:
 [1]  V1   V2   V24  V25  V26  V48  V63  V85  V102 V103 V144 V141 V140 
 [14] V224 V223 V222 V221 V294 V300

R> short_path_intensity$total_weight
[1] 0.1826603

R> short_path_cartheft <- ShortestPath(intnet_chicago, 
                                       node_id1 =  "V1" , 
                                       node_id2 =  "V300" , 
                                       weight =  "cartheft")
R> short_path_cartheft$path
+ 24/338 vertices, named, from d8a2c9e:
 [1]  V1   V2   V24  V25  V26  V48  V63  V85  V104 V103 V144 V141 V142 
 [14] V135 V136 V137 V224 V223 V225 V282 V292 V293 V301 V300

R> short_path_cartheft$total_weight
[1] 0
\end{verbatim}

Apart from the shortest path detection, the \texttt{PathTotalWeight()} function provided by the \textbf{intensitynet} package additionally allows the extraction of the weights of any given (not necessarily shortest) path. Requiring (i) the network as \texttt{intensitynet} object, (ii) a vector specifying the nodes along the path under selection, and (iii) an optional parameter with the type of weight to be computed as arguments, its use is illustrated in the following code. Note that if no weight type is provided, \texttt{PathTotalWeight()}  returns the total amount of edges in the path, otherwise, returns the total sum of the specified weight.

\begin{verbatim}
R> path <- c("V89",  "V92",  "V111", "V162",  "V164")
R> PathTotalWeight(intnet_chicago, path = path)
[1] 3

R> PathTotalWeight(intnet_chicago, path = path, weight =  "intensity")
[1] 0.03296913

PathTotalWeight(intnet_chicago, path = path, weight =  "robbery")
[1] 0.6666667
\end{verbatim}

Note that the path must be specified as an ordered list corresponding to the movement on the graph, i.e. a path with starting point $\mathbf{v}_{89}$ and destination $\mathbf{v}_{164}$ specified by  $\mathbf{v}_{89}\rightarrow \mathbf{v}_{92}\rightarrow \mathbf{v}_{111}\rightarrow \mathbf{v}_{162}\rightarrow \mathbf{v}_{164}$. The nodes on a path can either be stated by their names as strings or IDs as integers, or in a separate object as exemplified below.

\begin{verbatim}
R> path <- c(89, 92, 111, 162, 164)
R> PathTotalWeight(intnet_chicago, path = path)
[1] 3
\end{verbatim}

\subsection{Autocorrelation}\label{sec:autocorrelation}

To increase its applicability to different contexts and allow for a high level of flexibility within a unified framework, the \textbf{intensitynet} packages provide several correlation statistics in addition to the intensity function estimation including local and global versions of  Moran $I$ \citep{moran}, Geary's $C$ \citep{geary}, and Getis and Ords $G$ statistic \citep{geatisord}. While these global statistics help to quantify potential interrelations among the intensity functions over the complete network, their local counterparts quantify the contribution of the intensity functions for each element on the global one. As such, the local versions allow the identification of important hot- or coldspots on the network \citep[see][for detailed treatment and mathematical details]{Anselin2018}. 

Both global and local network correlation functions can directly be computed through the \texttt{NodeGeneralCorrelation()} and   \texttt{NodeLocalCorrelation()} functions, respectively, from the \texttt{intensitynet} object, each of which requires three arguments including (i) an \texttt{intensitynet} object, (ii)  the type of statistic (either \texttt{moran},  \texttt{getis}, or  \texttt{geary}), and (iii) a vector containing the nodewise intensities. Recalling the internal computation for different types of graphs, this vector could correspond to different objects, i.e.   \texttt{intensity\_in}, \texttt{intensity\_out}, \texttt{intensity\_und}, or  \texttt{intensity\_all}.
Using the adjacency matrix $\adj$ to specify the neighborhood structure, i.e. the   \emph{lags} over the network, to compute the spatial autocorrelations among the distinct entities, any correlation function is only available at node-level. Formally, the underlying \emph{lag structure} can be constructed for different orders, i.e. the $k$-order neighborhood, and includes either the partial or cumulative neighbors which derive directly from the corresponding adjacency matrices of and up to $\adj^k$ of order $k$, respectively. We note that the implemented approach can also be used to include bivariate  \citep{lee2001} or partial and semi-partial \citep{eckardtpartial} autocorrelation statistics in a  straightforward manner. 

The  \texttt{NodeLocalCorrelation()} functions return a list with the computational results stored as node attributes and an \texttt{intensitynet} object as arguments.  The name of the node attribute corresponds to the type argument used in the computation, i.e. \texttt{moran},  \texttt{getis}, or \texttt{geary}. For the local Moran $I$, the function returns a ($n\times 5$)-dimensional table as output 
which contains the empirical local Moran $I_i$ indices (\texttt{Ii}), the corresponding expected values $\mathds{E}\left[i_i\right]$ (\texttt{E.Ii}) and variances $\var\left[I_i\right]$ (\texttt{Var.Ii}), $Z$-value (\texttt{Z.Ii}) and a $p$-valued (\texttt{Pr(z != E(Ii))}) based on a Normal distribution. This table is augmented by an \texttt{intensitynet} object (only storing the permutation data ($\mathds{P}(z != \mathds{E}(I_i))$) from the node attributes). To illustrate its application, the next lines outline the computation of the \texttt{NodeLocalCorrelation()} using the \texttt{intnet\_chicago} as a toy example.

Initially, to extract the nodewise intensity functions from the node attributes of the \texttt{intensi\-tynet} object, the \texttt{vertex\_attr()} function from the \texttt{igraph} package can be applied to the graph object \texttt{intnet\_chicago\$graph}. In addition to the mandatory \texttt{intensitynet} object, this information can then be used as an argument of the \texttt{NodeLocalCorrelation()} function to compute the desired local autocorrelation statistic. Depending on the \texttt{type} argument provided by the user, the function either computes the local Moran $I$ (\texttt{moran}), Geary $C$ (\texttt{geary}), or  Getis $G$ (\texttt{getis}).

\begin{verbatim}
R> intensity_vec <- igraph::vertex_attr(intnet_chicago$graph)$intensity

R> data_moran <- NodeLocalCorrelation(intnet_chicago, 
+                                     dep_type =  "moran" , 
+                                     intensity = intensity_vec)
R> head(data_moran)
            Ii          E.Ii     Var.Ii       Z.Ii Pr(z != E(Ii))
V1  0.31576139 -0.0009369774 0.31640163  0.5630234      0.5734189
V2  0.12649811 -0.0009369774 0.07839415  0.4551423      0.6490069
V3  0.31576139 -0.0009369774 0.31640163  0.5630234      0.5734189
V4 -0.44129172 -0.0009369774 0.31640163 -0.7828586      0.4337102
V5 -0.04934006 -0.0018300486 0.15297793 -0.1214705      0.9033184
V6 -0.44129172 -0.0009369774 0.31640163 -0.7828586      0.4337102

R> intnet_chicago <- data_moran$intnet

R> data_geary <- NodeLocalCorrelation(intnet_chicago, 
+                                     dep_type =  "geary" , 
+                                     intensity = intensity_vec)
R> head(data_geary$correlation)
[1] 0.0000000 0.4524253 0.0000000 1.8097012 1.0263252 1.8097012

R> intnet_chicago <- data_geary$intnet
R> data_getis <- NodeLocalCorrelation(intnet_chicago, 
+                                     dep_type =  "getis" , 
+                                     intensity = intensity_vec)
R> head(data_getis$correlation)
[1] -0.5630234 -0.4551423 -0.5630234  0.7828586 -0.1214705  0.7828586

R> intnet_chicago <- data_getis$intnet
\end{verbatim}


The correlation and covariance among the nodewise intensities, and the global Moran's $I$ and Geary's $C$ autocorrelation statistics provided by the \texttt{NodeGeneral\-Correlation()} function are constructed as wrapper functions using the \texttt{nacf()} function from the package \texttt{sna} \citep{butts2008} as the source. Internally, this function transforms the \texttt{intensitynet} object into a \texttt{sna} format to meet the desired input of the \texttt{nacf()} function. In addition to \texttt{intensitynet} object, the implemented function also requires (i) a type statement (\texttt{correlation},  \texttt{covariance},  \texttt{moran}  or  \texttt{geary}), (ii) a vector constructed from the nodewise intensities and (iii) a number specifying the \emph{order} of $\adj^k$. Apart from these mandatory arguments, the user could also specify if a partial neighborhood (default) or a cumulative lag structure up to order $k$ should be used in the computations. The partial neighborhood for any given node $v_i$ is defined as the set of nodes that are exactly $k$ steps apart from $v_i$ whereas the cumulative neighborhood of $v_i$ subsumes all nodes whose distance to $v_i$ is less than or equal to $k$. As output, the corresponding results are provided in form of a vector of length $k$, starting with zero neighbors where no lag information is used to the $k$-th neighbors as specified in the \texttt{order} statement. An application of the \texttt{NodeGeneralCorrelation()} function to the \texttt{intnet\_chicago} is illustrated in the following lines.

Using the \texttt{intensity\_vec} object from the previous example as input for our computations, execution of the following lines yields the partial covariances of the nodewise intensity functions for the sets of partial neighbors of orders $0$ to $2$.

\begin{verbatim}
R> NodeGeneralCorrelation(intnet_chicago, 
+                         dep_type =  "covariance" , 
+                         lag_max = 2, 
+                         intensity = intensity_vec)
           0            1            2 
3.742840e-05 1.746413e-05 8.142177e-06 
\end{verbatim}

Likewise, the partial correlation for the sets of partial neighbors up to order $k=5$  and its cumulative counterpart version (\texttt{partial\_neighborhood= FALSE}) can be obtained by running the following code. Both quantities show a clear decrease in correlation from lag zero, corresponding to the correlation of the calculated nodewise intensity values with themselves, to lag five.

\begin{verbatim}
R>  NodeGeneralCorrelation(intnet_chicago, 
+                          dep_type =  "correlation" , 
+                          lag_max = 5, 
+                          intensity = intensity_vec)
         0          1          2          3          4          5 
1.00000000 0.46522061 0.21689644 0.18078130 0.09940217 0.03468309 

R>  NodeGeneralCorrelation(intnet_chicago, 
+                          dep_type =  "correlation" , 
+                          lag_max = 5, 
+                          intensity = intensity_vec,
+                          partial_neighborhood = FALSE)
        0         1         2         3         4         5 
1.0000000 0.4652206 0.3012361 0.2402657 0.1827256 0.1317444 
\end{verbatim}

\subsection{Data Visualization}

Working with event data on relational systems, graphical outputs are of particular importance to describe the observed data and computed results. To allow for simple computations, different visualization tools are included in the \textbf{intensitynet} package. All main visual functionalities are provided by the \texttt{PlotHeatmap()} function which serves as a generic visualization tool and allows for the computation of graphically optimized heatmap representations of the network, either with or without any additional information on the event locations or estimated quantities. The function can directly be applied to any \texttt{intensitynet} object while the output can be specified through a wide range of additional arguments. The \texttt{PlotHeatmap()} function internally uses the  \texttt{GeoreferencedGgplot2()} function from the \texttt{netTools} class which, in turn, is embedded into the \textbf{ggplot2} framework. As such, the function also allows for further arguments to manipulate the plot by customizing e.g. the edge or node size and color layers.

\begin{figure}[htb]
	\begin{center}
		\scalebox{0.8} {\includegraphics{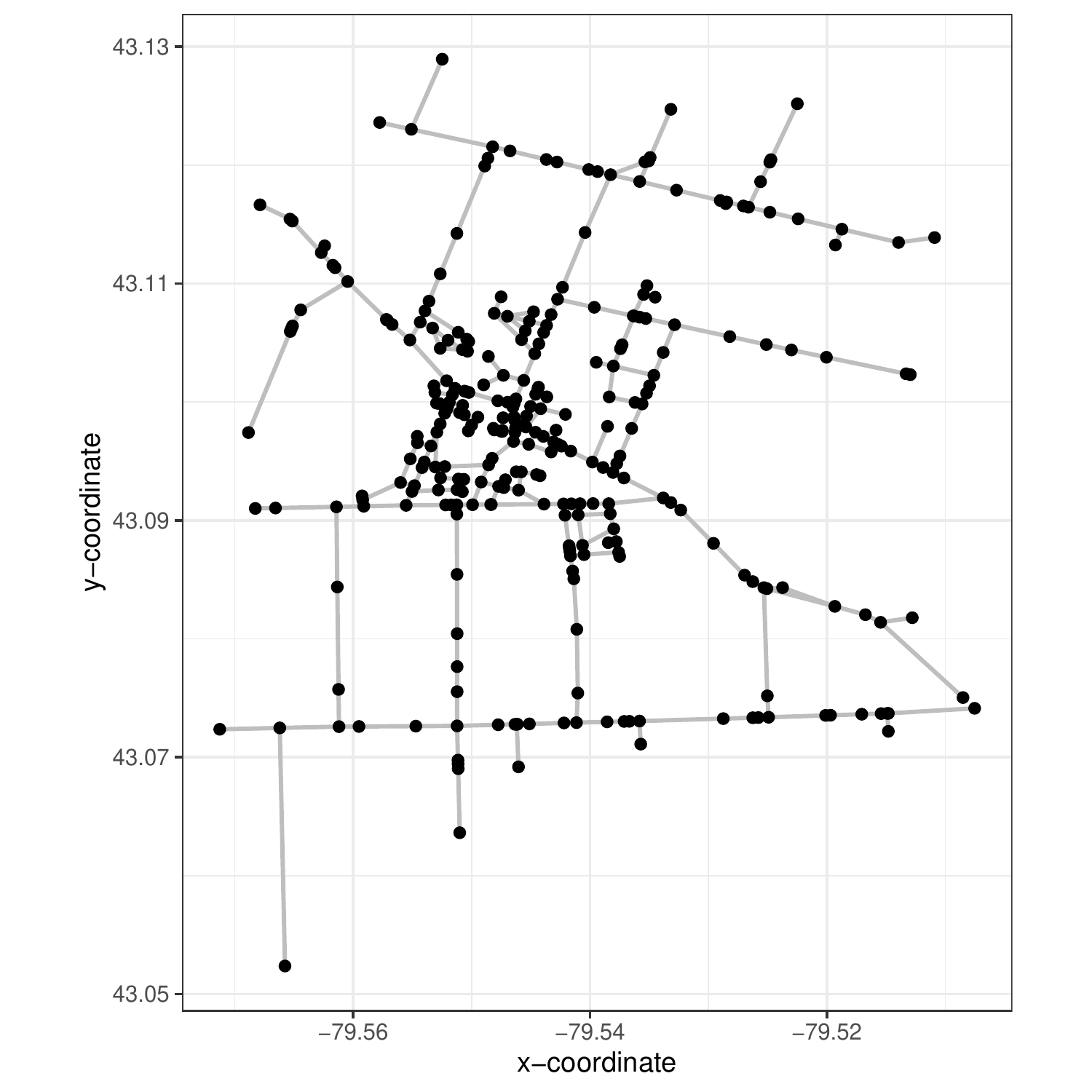}}
	\end{center}
	\caption{Imported network structured of the Ontario data example from section \ref{sec:shapefile} which nodes represented as black dots and edges shown as solid gray lines}
	\label{fig:ontario:example}
\end{figure}

Without any further arguments, \texttt{PlotHeatmap()} yields a graphical output of the internally stored network structure including all edges and vertices (see Figure \ref{fig:ontario:example} for a visual representation of the underlying graph structure for the \texttt{intnet\_orn} object presented in section \ref{sec:shapefile}).
In addition to the basic visualization of the network structure itself, the precise event locations on the individual edges can be added to the output by setting the parameter \texttt{show\_events} to \texttt{TRUE}. As a result, each event location is shown as orange squares. The transparency of the squares can be further customized by using the  \texttt{alpha}  argument. 

\begin{figure}[h]
	\begin{center}
		\begin{tabular}{cc}
    		\scalebox{0.45}[0.335] {\includegraphics{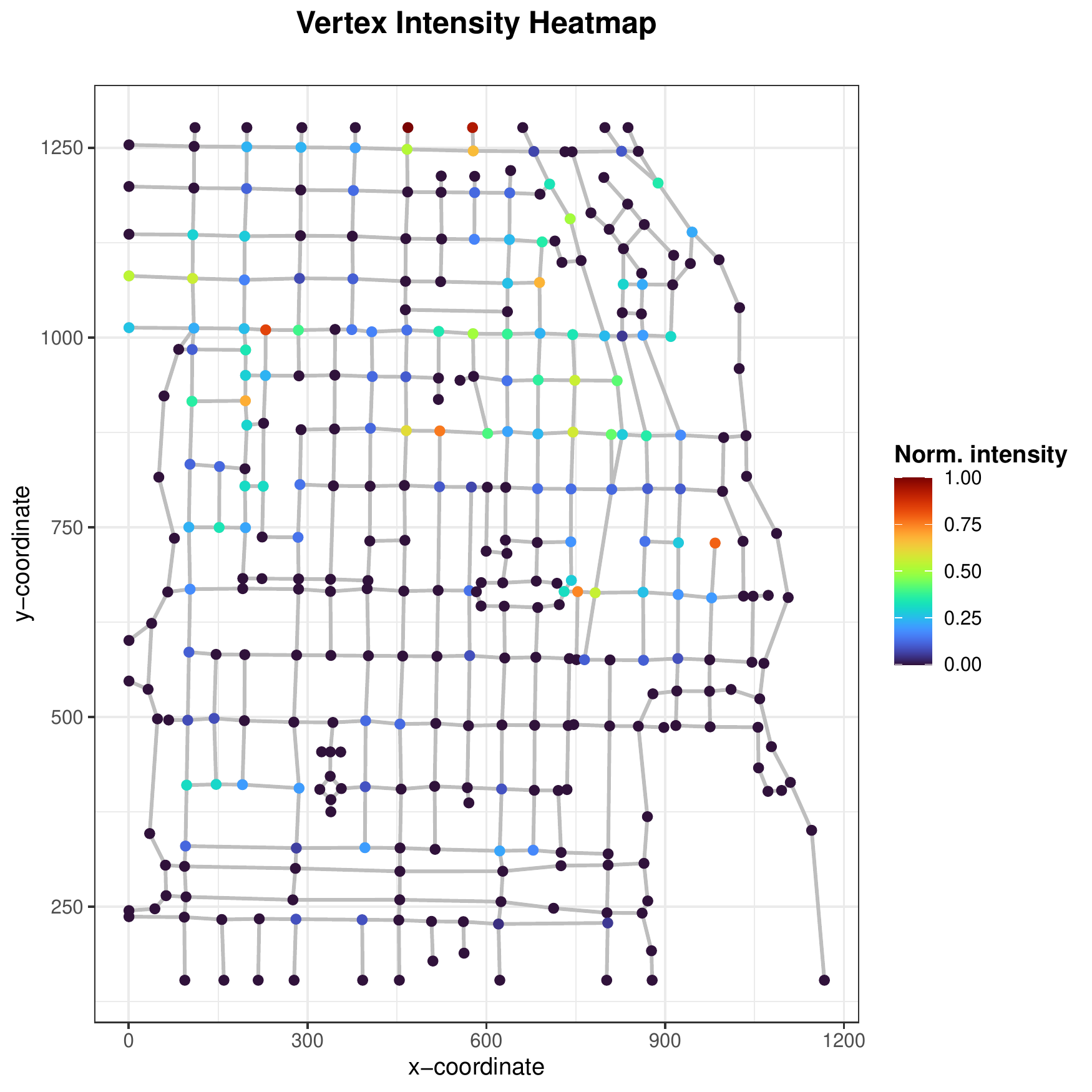}} &
    		\scalebox{0.45}[0.33]{\includegraphics{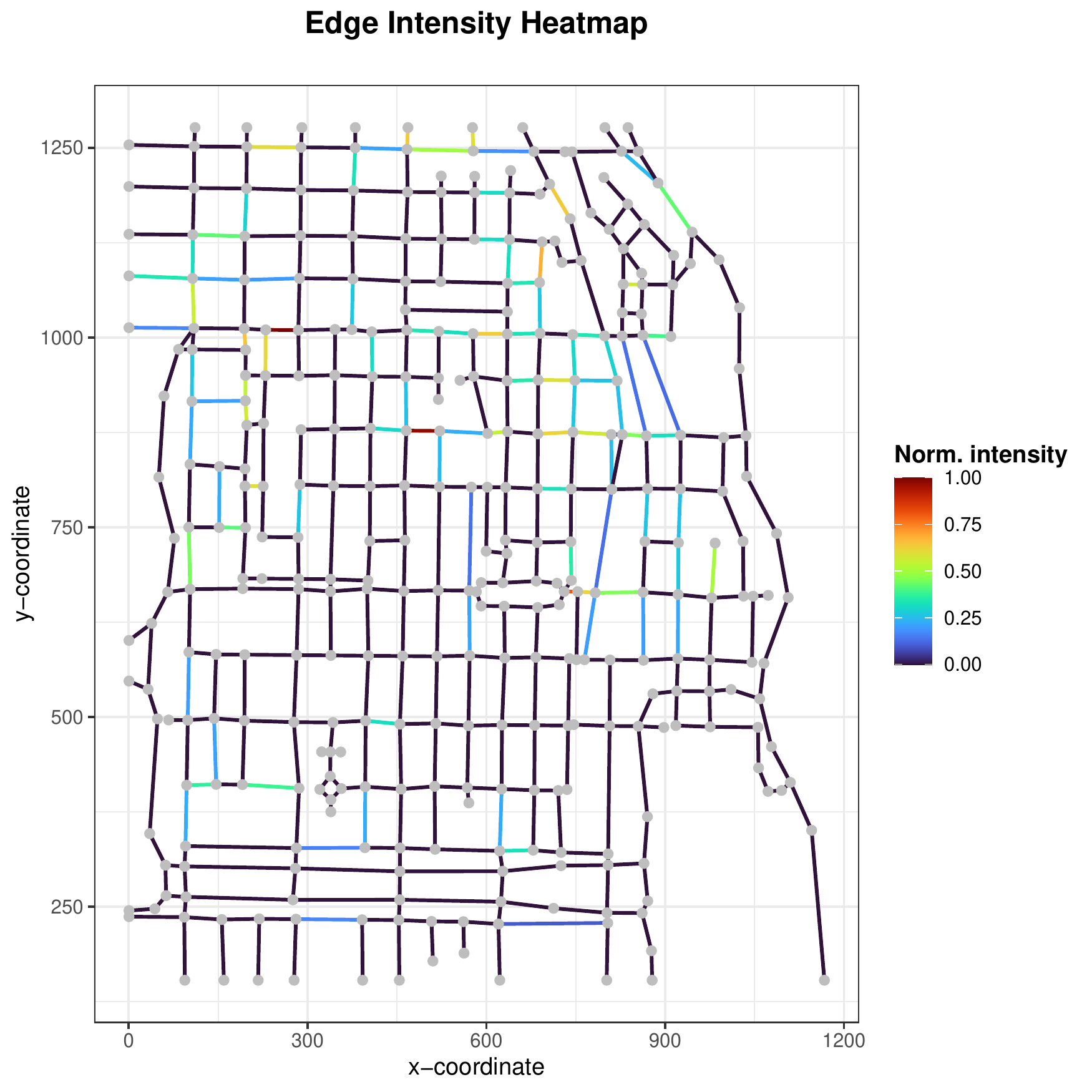}} \\
    		
    		\scalebox{0.4}[0.335]{\includegraphics{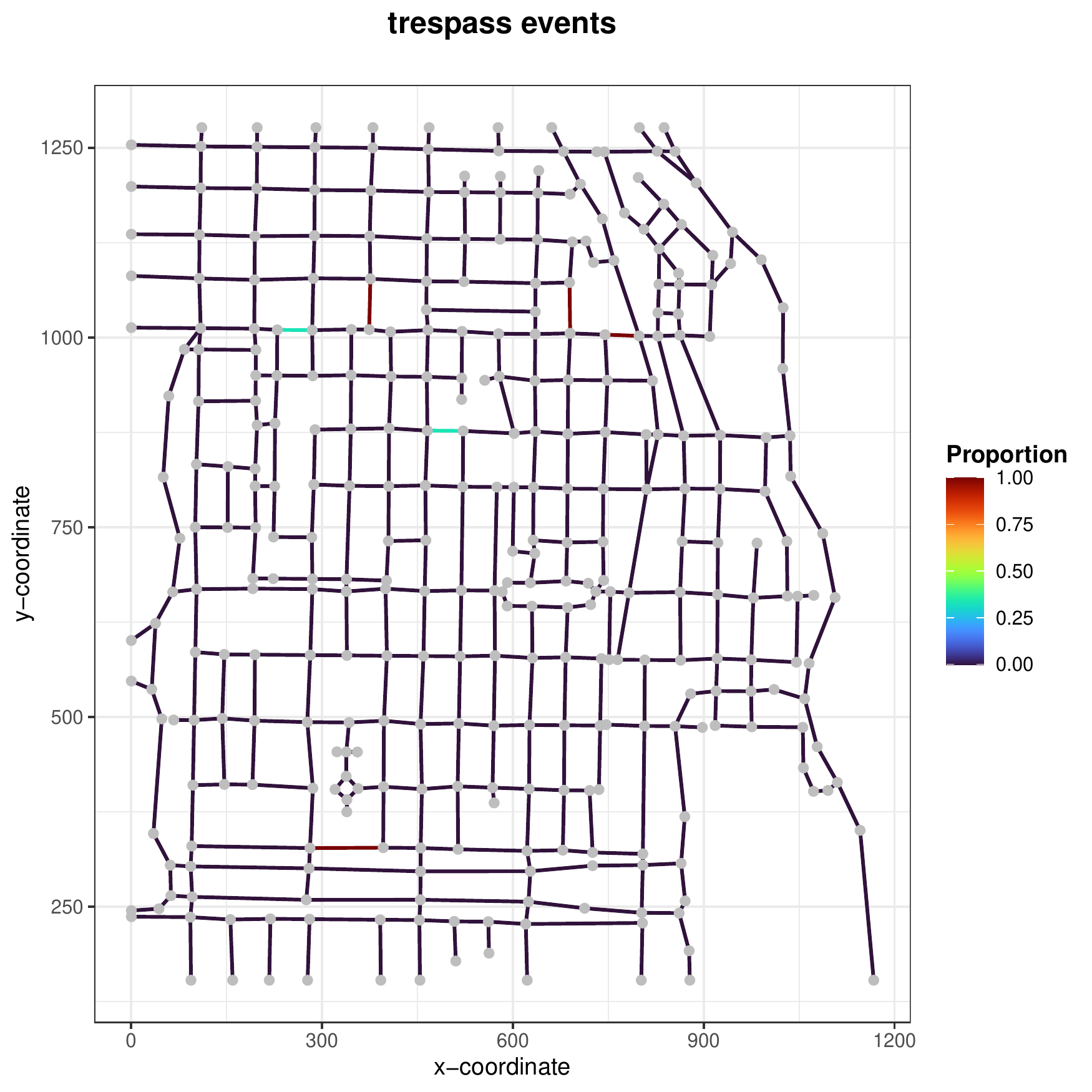}} & 
    		\scalebox{0.4}[0.335]{\includegraphics{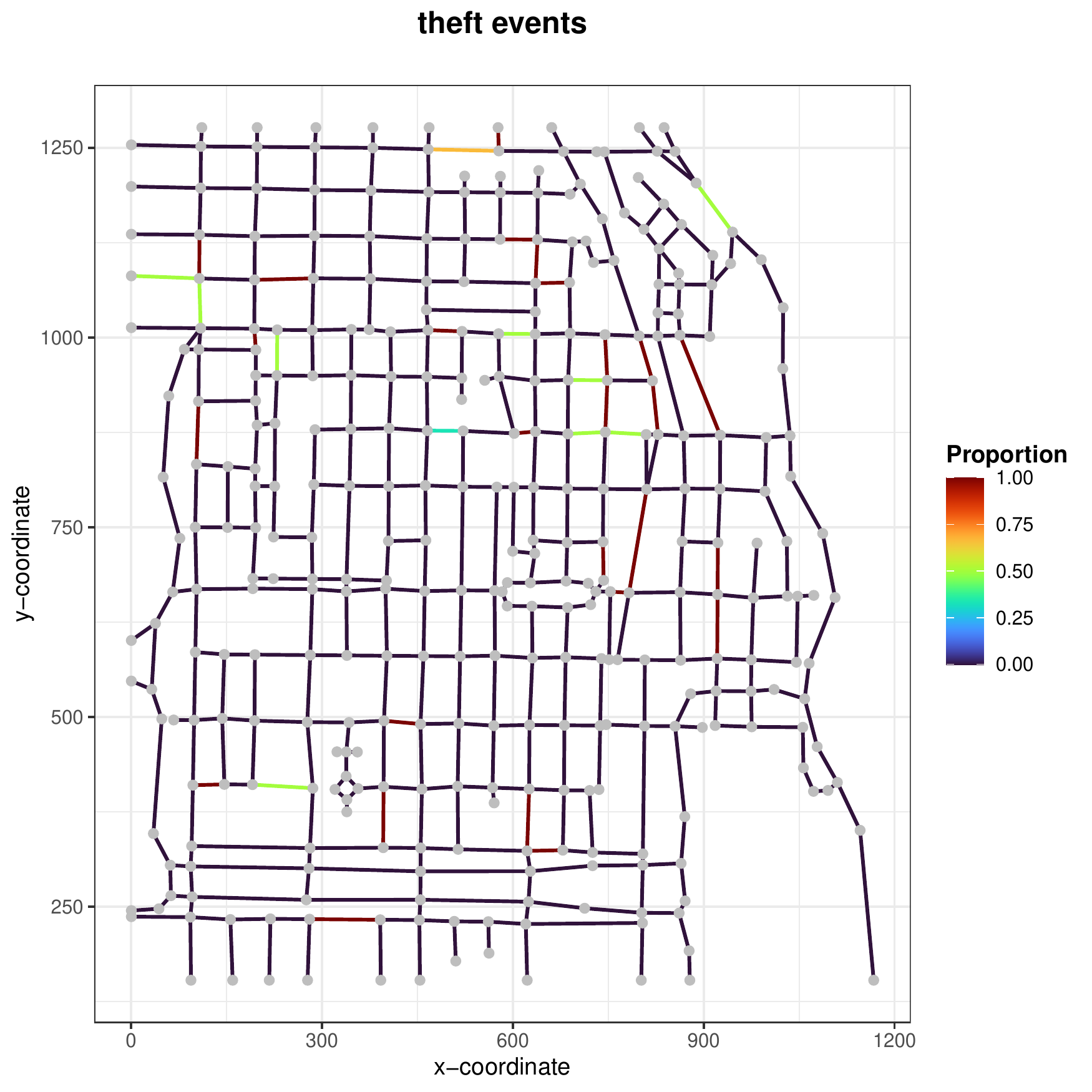}} \\ 
    			
    		\scalebox{0.44}[0.33] {\includegraphics{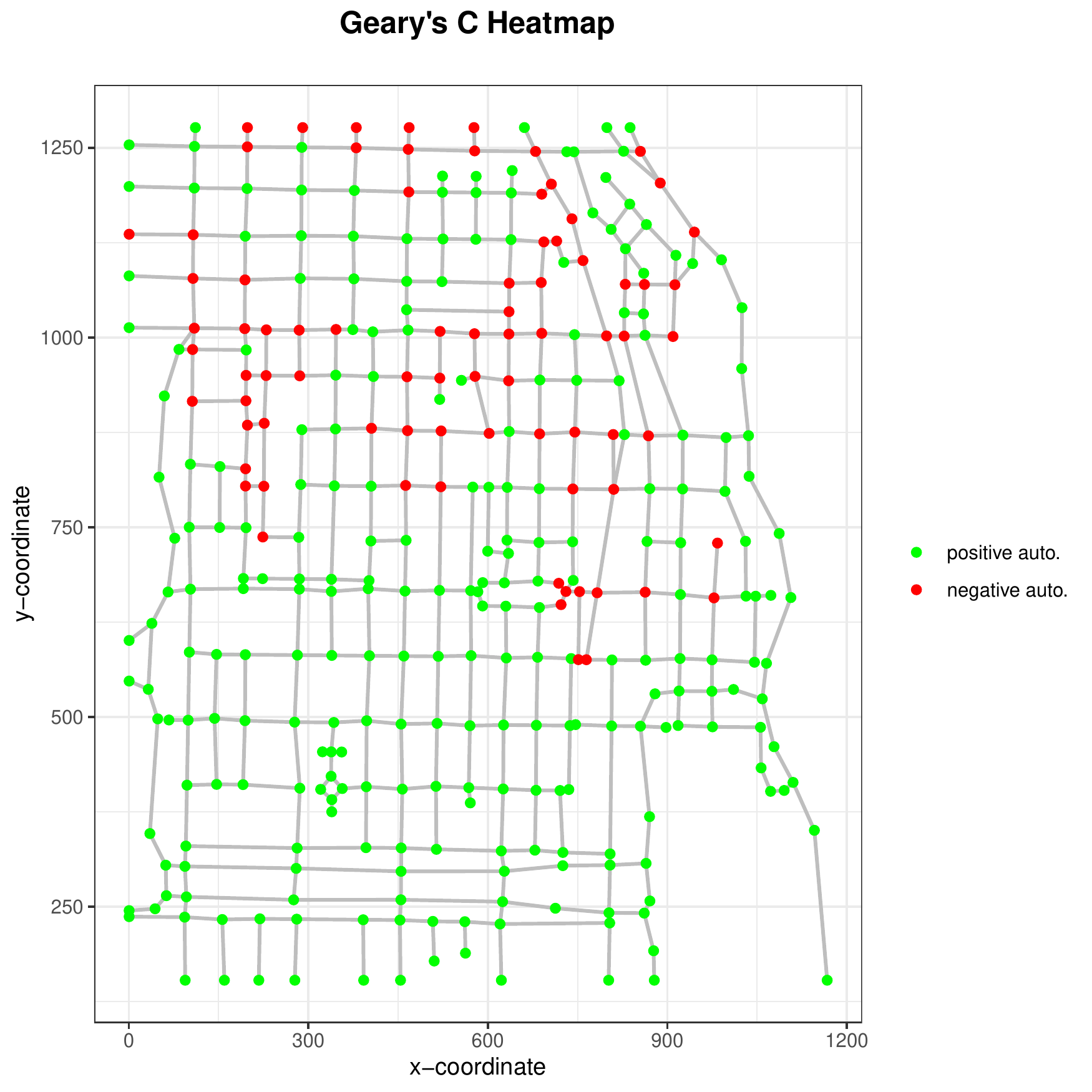}} &
    		\scalebox{0.47}[0.335]{\includegraphics{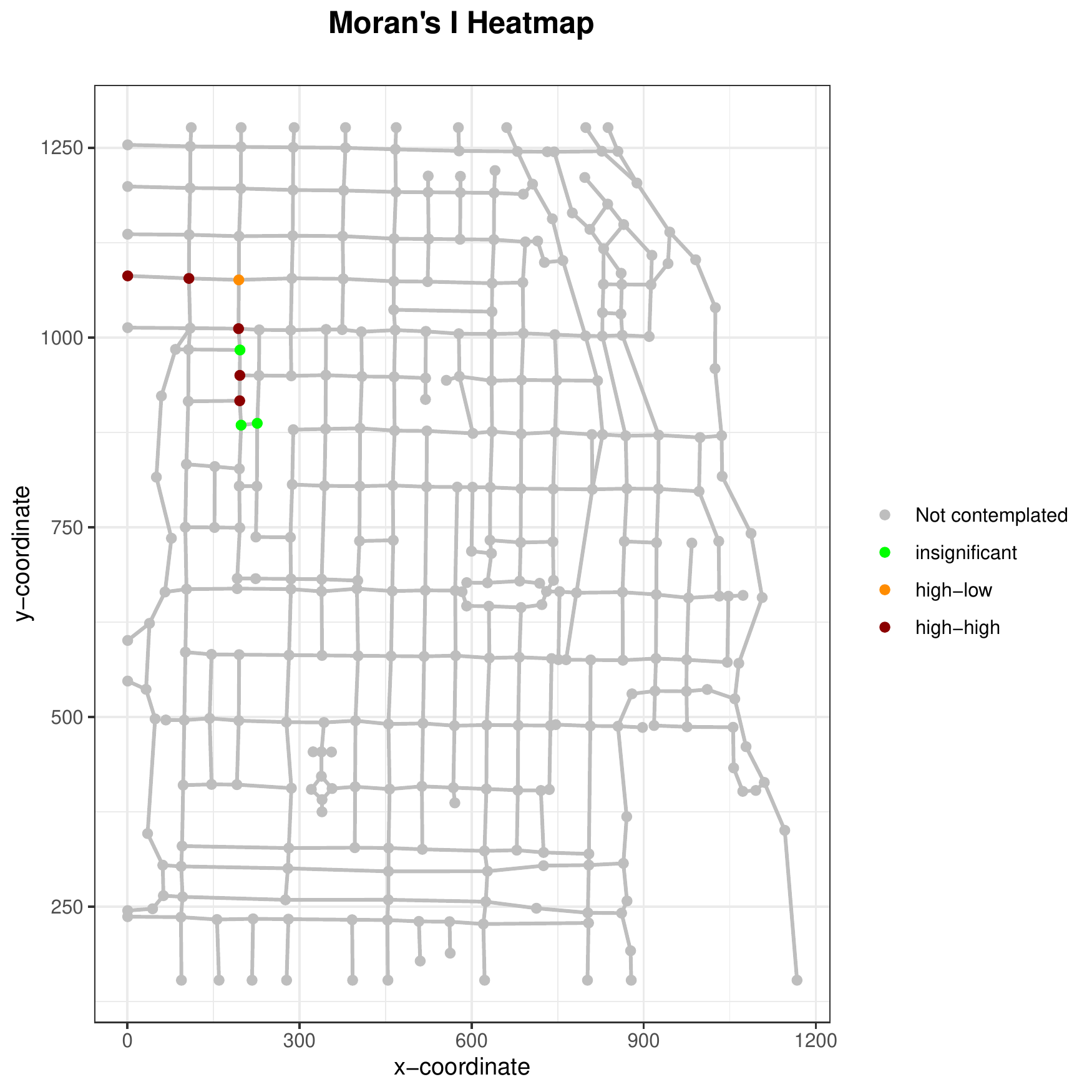}} 
		\end{tabular}
	\end{center}
	\caption{Heatmap representations of network intensity functions and derived characteristics computed from the \texttt{chicago} data with the rows defined as follows: The top row depicts the heatmaps for the nodewise (left) and edgewise intensity functions (right), the central row consists of the heatmaps constructed from the edgewise proportions of the crime categories \textit{trespass}  (left) and  \textit{theft}  (right), and the bottom panels show the heatmap representations of the local Geary's $C$ (left) and Moran's $I$ (right) autocorrelation statistics.  Positive and negative autocorrelations are highlighted in both Geary's $C$ and Morans $I$ heatmaps.}
	\label{fig:heatmap:examples}
\end{figure}

Apart from the raw network and event information, the \texttt{PlotHeatmap()}  function also allows for the representation of different results derived from the original input information including various local correlation functions, categorical marked point proportions or averages, and vertex or edge intensities as outlined in Sections \ref{sec:data:manipulation} to \ref{sec:autocorrelation}. Any such intensity-based characteristics can be included in the output by specifying the \texttt{heat\_type} argument. Examples of heatmap representations of the edgewise (\texttt{heat\_type =  e\_intensity}) nodewise (\texttt{heat\_type =  v\_intensity}) intensity functions, the mark proportions (\texttt{heat\_type =  trespass}), and Geary's $C$ (\texttt{heat\_type =  geary}) are depicted in Figure \ref{fig:heatmap:examples}.

In addition to the global representation of the network and all event-based computations over the complete relational system under study, the \texttt{PlotHeatmap()} function allows also the computation of graphical outputs for only some pre-selected vertices or edges. To construct a visualization of the localized network only, the selected vertices or edges need to be specified in advance either by using the vertex identifier or an \texttt{igraph} vertices or edges object, i.e. \texttt{igraph.vs} or \texttt{igraph.es}, respectively. The following lines of code outline the localized representation of the local Moran's $I$ autocorrelation statistic for a list of pre-specified vertices (V66,  V65,  V64,  V84,  V98,  V101,  V116,  V117,  V118). The corresponding output is shown in the right bottom of Figure \ref{fig:heatmap:examples}).

\begin{verbatim}
R> PlotHeatmap(intnet_chicago, 
+              heat_type =  "moran" , 
+              net_vertices = c("V66",  "V65",  "V64", 
+                               "V84",  "V98",  "V101", 
+                               "V116",  "V117", "V118")) 
\end{verbatim}



In addition to the \texttt{PlotHeatmap()} function, the \textbf{intensitynet} package also provides a  \texttt{plot()} function which is internally designed as a wrapper function of the \texttt{Georeferenced\-Plot()} function provided in the auxiliary main class \texttt{netTools} and allows to uses the capabilities of the \texttt{plot.igraph()} function from the \textbf{igraph} package. As such, the \texttt{plot()} function allows to modify the nodes and edge labels of the network output and to highlight movements along the networks, i.e. paths, by specifying the corresponding vertex identifiers. Moreover, this function also allows for a proper spatial embedding of the network and the plotting of background grids, corresponding coordinates on the axes, and highlighting of event locations using simple flag arguments (\texttt{TRUE}, \texttt{FALSE}). As in the \texttt{PlotHeatmap()} function, the transparency of the events can be determined by using the \texttt{alpha} parameter. Finally, using the capacities of the \texttt{plot.igraph()} function, the \texttt{plot()} function allows to modify the output by including any graph analytic quantity from the \textbf{igraph} toolbox including communities or centrality measure based weights. Its applications to the \texttt{intnet\_chicago} object is illustrated in the next lines of code.

\begin{verbatim}
R> plot(intnet_chicago, show_events = TRUE)
R> short_path <- ShortestPath(intnet_chicago, 
+                             node_id1 =  "V1" , 
+                             node_id2 =  "V300" , 
+                             weight =  "intensity")
R> plot(intnet_chicago, show_events = TRUE, path = short_path$path)
\end{verbatim}

\begin{figure}[htb]
	\begin{center}
		\begin{tabular}{cc}
			\scalebox{0.4} {\includegraphics{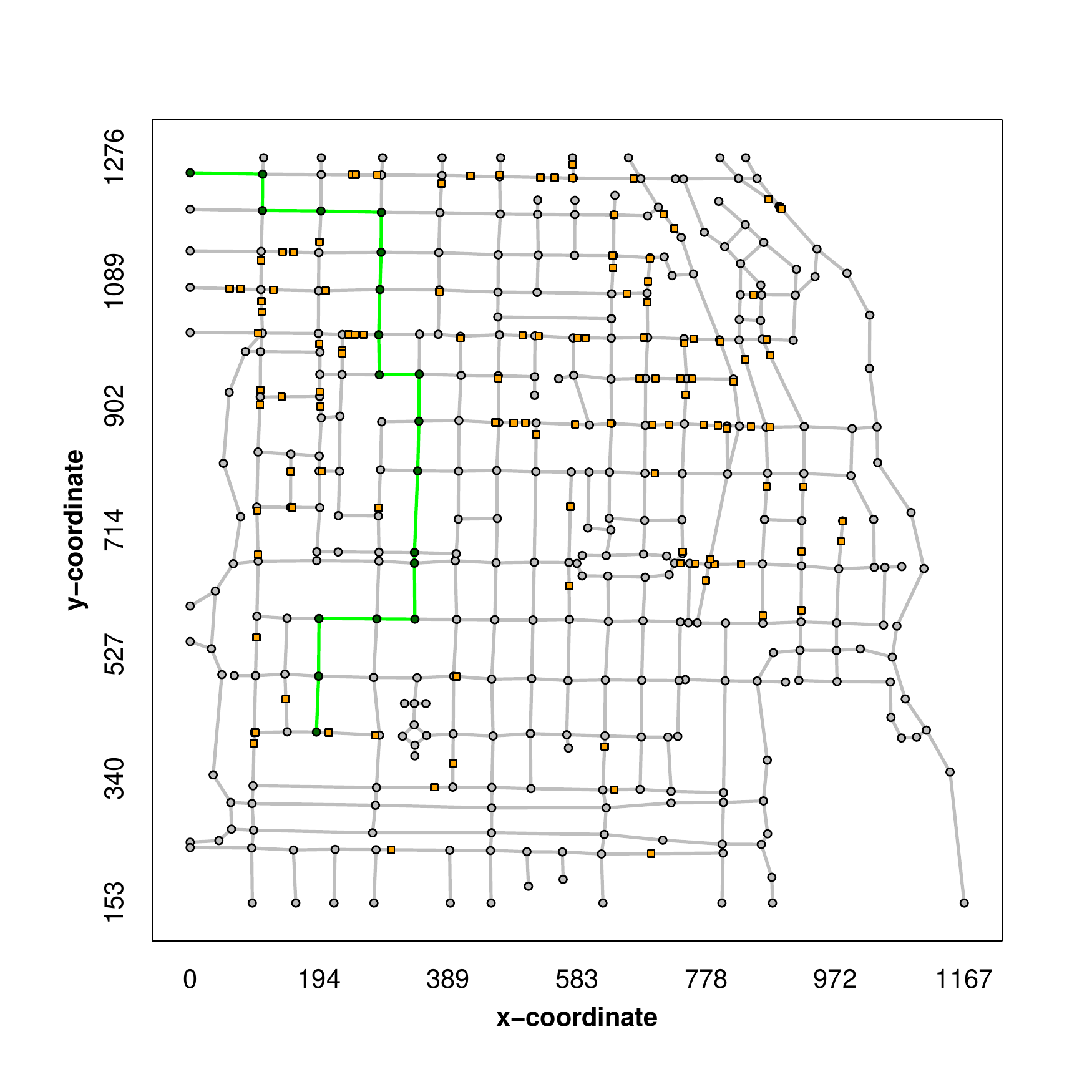}} & 
			\scalebox{0.4} {\includegraphics{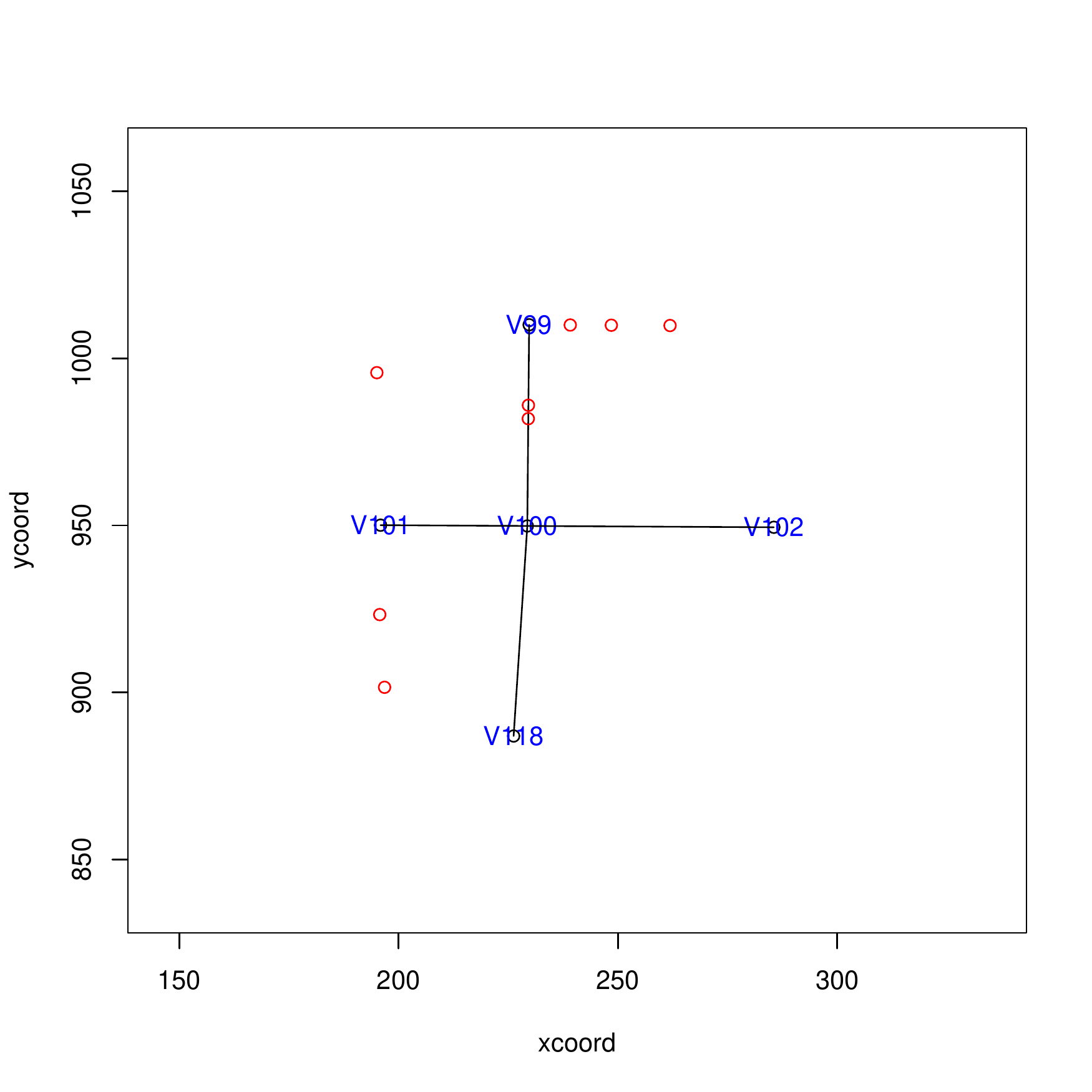}}
		\end{tabular}
	\end{center}
	\caption{Selected outputs of the \texttt{plot()} and \texttt{PlotNeighborhood()} functions computed from the \texttt{intnet\_chicago} object. Shortest path from $v_1$ to $v_{300}$ along the network which avoids criminal areas  highlighted in the color green (left), first-order neighborhood and incident edges for node $v_{100}$ and closest raw event locations (red circles).}
	\label{fig:shortpath:and:neighborhood}
\end{figure}

Both the \texttt{PlotHeatmap()} and the \texttt{plot()} functions are augmented by the  \texttt{PlotNeighborhood()} function which allows to highlight the first-order neighborhood for user-specified nodes and related event locations (highlighted by red circles). Execution of the code below yields the corresponding results for node $v_{100}$ ($V100$), the resulting plot from the execution of this code is shown in Figure \ref{fig:shortpath:and:neighborhood} on the right.

\begin{verbatim}
R> PlotNeighborhood(intnet_chicago, node_id =  "V100")
\end{verbatim}

\FloatBarrier

\subsection{Going beyond points}

Embedded into a graph theoretic formulation, the \texttt{intensitynet} package allows for additional investigations of the network event patterns that go well beyond the classic approaches in spatial point process analysis. The network assigned event locations or computed network intensity functions allow for the specification of advanced spatial (spatio-temporal) regression models for network-based responses and additional spatial covariates or mark information \citep[see][for recent regression model for network-based response]{SNR}. Apart from regression specifications, the computed results can also be used to investigate the dependence or variation of different types of events, i.e. crimes, or to embed the observed network event patterns as graph-valued data into the framework of object-oriented data analysis. Finally, the inherent formulation of the network characteristics within the \textbf{igraph} framework allows using the well-established methodological toolbox and mathematical concepts of applied graph theory to detect communities and clusters, compute centrality and connectivity measures, and apply spectral analysis methods. 

Using the \texttt{chicago} data as a toy example, the following code investigates the potential correlation between the nodewise intensities of trespass and robbery events. 

\begin{verbatim}
R> chicago_trespass <- chicago_df[chicago_df$marks ==  "trespass" ,]
R> trespass_intnet <- intensitynet(adj_mtx, 
+                                  node_coords = node_coords, 
+                                  event_data = chicago_trespass)
R> trespass_intnet <- RelateEventsToNetwork(trespass_intnet)
Calculating edge intensities with event error distance of 5...
    |=============================================================| 100%

Calculating node intensities...
    |=============================================================| 100%

R> chicago_robbery <- chicago_df[chicago_df$marks ==  "robbery" ,]
R> robbery_intnet <- intensitynet(chicago_adj_mtx, 
+                                 node_coords = chicago_node_coords, 
+                                 event_data = chicago_robbery)
R> robbery_intnet <- RelateEventsToNetwork(robbery_intnet)
Calculating edge intensities with event error distance of 5...
  |=============================================================| 100%

Calculating node intensities...
  |=============================================================| 100%
R> trespass_intensity <- igraph::vertex_attr(trespass_intnet$graph, "intensity")
R> robbery_intensity <- igraph::vertex_attr(robbery_intnet$graph, "intensity")
R > cor(trespass_intensity, robbery_intensity)
[1] 0.3559283
\end{verbatim}

Extracting the \texttt{igraph} network structure from an \texttt{intensitynet} object, the next lines illustrate the computation of graph theoretic quantities from the \texttt{intnet\_chicago} object.

\begin{verbatim}
R> g <- intnet_chicago$graph

R> degree <- igraph::degree(g)
R> head(degree)
V1 V2 V3 V4 V5 V6 
 1  4  1  1  4  1 
R> max(degree)
[1] 5

R> head(igraph::betweenness(g))
    V1     V2     V3     V4     V5     V6 
  0.00 677.50   0.00   0.00 797.25   0.00 

R> head(igraph::edge_betweenness(g))
[1] 337.00 337.00 461.75 556.25 337.00 527.00

R> head(igraph::closeness(g))
          V1           V2           V3           V4           V5           V6 
2.850374e-06 3.183678e-06 3.101290e-06 3.303551e-06 3.399421e-06 3.481001e-06 
\end{verbatim}

%

%

%

\section{Summary and discussion} \label{sec:summary}
This paper has introduced the general implementation and main functionalities of the \textbf{intensitynet} package including exploratory analysis tools, network adjusted summary and autocorrelation statistics, and visualization capacities. Bridging the general ideas from graph theory into the analysis of structured spatial event-type data, the package allows for the analysis of potential events on different types or combinations of edges and also disconnected subgraph structures. The package is build around a set of generic intensities functions which derive directly from edgewise computations and allow to address different graph theoretic entities and sets in the network under study and help to investigate the variation in numbers of events over the graph object. Embedded into the  \textbf{igraph} framework, the \textbf{intensitynet}  sets the ground for even further analysis based on graph theory methodologies and also easy importation into related \texttt{R} packages.


Although the package is designed to work with any network size and edge configuration, both the size of the network and the number of events under study affect the computing time. To address the computational burden and structure the necessary computational operations, the package is  implemented in the two separate functions (\texttt{intensitynet()} and \texttt{RelateEventsToNetwork()}). This enables the user to decide to either obtain some intensity based statistics or work explicitly with the network representation in the form of an \textbf{igraph} class object.\\
To test the reliability of our package, we ran numerous unitary tests, and also general tests, using multiple datasets, some of them with dummy data and some involving real datasets such as the Chicago data presented in this work. Additionally, the package has passed all the CRAN repository policies and checks. Moreover, the package is in constant revision and development, to detect and amend unnoticed bugs and other possible code-related problems. 

 About future development, some of the features that could be implemented are (1) consider more input options to initialize the \texttt{intensitynet()}, (2) incorporate an occurrence legend in the plots \texttt{plot()} or \texttt{PlotHeatmap()} if the occurrences are chosen to be shown, (3) consider multiple output options to save the network model in different formats, and finally (4) develop an application to access interactively all the tools that our package offers, using the \textbf{Shiny} package.

\section*{Acknowledgments}
The authors gratefully acknowledge the financial support of the Barcelona economic association \textit{Amics del Pa\'{i}s} and the support of the MCIN/AEI/ 10.13039/501100011033 (Spanish Government) through Grant PID2020-115442RB-I00,  Matthias Eckardt was funded by the Walter Benjamin grant 467634837 from the German Research Foundation.


\bibliographystyle{ecta}
\bibliography{refs}

\begin{thebibliography}{36}
\newcommand{\enquote}[1]{``#1''}
\expandafter\ifx\csname natexlab\endcsname\relax\def\natexlab#1{#1}\fi

\bibitem[\protect\citeauthoryear{Anderes, M{\o}ller, and Rasmussen}{Anderes
  et~al.}{2020}]{10.1214/19-AOS1896}
\textsc{Anderes, E., J.~M{\o}ller, and J.~G. Rasmussen} (2020):
  \enquote{{Isotropic covariance functions on graphs and their edges},}
  \emph{The Annals of Statistics}, 48, 2478 -- 2503.

\bibitem[\protect\citeauthoryear{Ang, Baddeley, and Nair}{Ang
  et~al.}{2012}]{10.2307/23357213}
\textsc{Ang, Q., A.~Baddeley, and G.~Nair} (2012): \enquote{Geometrically
  Corrected Second Order Analysis of Events on a Linear Network, with
  Applications to Ecology and Criminology,} \emph{Scandinavian Journal of
  Statistics}, 39, 591--617.

\bibitem[\protect\citeauthoryear{Anselin}{Anselin}{2018}]{Anselin2018}
\textsc{Anselin, L.} (2018): \enquote{A Local Indicator of Multivariate Spatial
  Association: Extending Geary's C,} \emph{Geographical Analysis}, 51,
  133--150.

\bibitem[\protect\citeauthoryear{Baddeley, Nair, Rakshit, McSwiggan, and
  Davies}{Baddeley et~al.}{2021}]{BADDELEY2021100435}
\textsc{Baddeley, A., G.~Nair, S.~Rakshit, G.~McSwiggan, and T.~M. Davies}
  (2021): \enquote{Analysing point patterns on networks — A review,}
  \emph{Spatial Statistics}, 42, 100435.

\bibitem[\protect\citeauthoryear{Baddeley, Rubak, and Turner}{Baddeley
  et~al.}{2015}]{spatstat:book}
\textsc{Baddeley, A., E.~Rubak, and R.~Turner} (2015): \emph{Spatial Point
  Patterns: Methodology and Applications with {R}}, Chapman and Hall/CRC Press.

\bibitem[\protect\citeauthoryear{Baddeley and Turner}{Baddeley and
  Turner}{2005}]{spatstat:JSS}
\textsc{Baddeley, A. and R.~Turner} (2005): \enquote{{spatstat}: An {R} Package
  for Analyzing Spatial Point Patterns,} \emph{Journal of Statistical
  Software}, 12, 1--42.

\bibitem[\protect\citeauthoryear{Bivand and Lewin-Koh}{Bivand and
  Lewin-Koh}{2019}]{maptools}
\textsc{Bivand, R. and N.~Lewin-Koh} (2019): \emph{maptools: Tools for Handling
  Spatial Objects}, r package version 0.9-9.

\bibitem[\protect\citeauthoryear{Briz-Redon}{Briz-Redon}{2021{\natexlab{a}}}]{DRHotNet}
\textsc{Briz-Redon, A.} (2021{\natexlab{a}}): \emph{DRHotNet: Differential Risk
  Hotspots in a Linear Network}, r package version 2.0.

\bibitem[\protect\citeauthoryear{Briz-Redon}{Briz-Redon}{2021{\natexlab{b}}}]{SpNetPrep}
---\hspace{-.1pt}---\hspace{-.1pt}--- (2021{\natexlab{b}}): \emph{SpNetPrep:
  Linear Network Preprocessing for Spatial Statistics}, r package version 1.2.

\bibitem[\protect\citeauthoryear{Butts}{Butts}{2008}]{butts2008}
\textsc{Butts, C.~T.} (2008): \enquote{Social Network Analysis with sna,}
  \emph{Journal of Statistical Software}, 6, 1--51.

\bibitem[\protect\citeauthoryear{Csardi and Nepusz}{Csardi and
  Nepusz}{2006}]{igraph}
\textsc{Csardi, G. and T.~Nepusz} (2006): \enquote{The igraph software package
  for complex network research,} \emph{InterJournal}, Complex Systems, 1695.

\bibitem[\protect\citeauthoryear{Diestel}{Diestel}{2016}]{graphtheory}
\textsc{Diestel, R.} (2016): \emph{Graph Theory}, Springer.

\bibitem[\protect\citeauthoryear{Eckadt, Klein, Greven, and Mateu}{Eckadt
  et~al.}{2022}]{SNR}
\textsc{Eckadt, M., N.~Klein, S.~Greven, and J.~Mateu} (2022):
  \enquote{Structured additive distributional regression for spatial point
  patterns on network domains,} \emph{Working paper}.

\bibitem[\protect\citeauthoryear{Eckardt and Mateu}{Eckardt and
  Mateu}{2018}]{Eckardt2018}
\textsc{Eckardt, M. and J.~Mateu} (2018): \enquote{Point Patterns Occurring on
  Complex Structures in Space and Space-Time: An Alternative Network Approach,}
  \emph{Journal of Computational and Graphical Statistics}, 27, 312--322.

\bibitem[\protect\citeauthoryear{Eckardt and Mateu}{Eckardt and
  Mateu}{2021{\natexlab{a}}}]{eckardtpartial}
---\hspace{-.1pt}---\hspace{-.1pt}--- (2021{\natexlab{a}}): \enquote{Partial
  and Semi-Partial Statistics of Spatial Associations for Multivariate Areal
  Data,} \emph{Geographical Analysis}, 53, 818--835.

\bibitem[\protect\citeauthoryear{Eckardt and Mateu}{Eckardt and
  Mateu}{2021{\natexlab{b}}}]{Eckardt2021}
---\hspace{-.1pt}---\hspace{-.1pt}--- (2021{\natexlab{b}}):
  \enquote{Second-order and local characteristics of network intensity
  functions,} \emph{TEST}, 30, 318--340.

\bibitem[\protect\citeauthoryear{Fox and Bolton}{Fox and
  Bolton}{2002}]{engineermaths}
\textsc{Fox, H. and W.~Bolton} (2002): \emph{Mathematics for Engineers and
  Technologists}, Butterworth-Heinemann.

\bibitem[\protect\citeauthoryear{Geary}{Geary}{1954}]{geary}
\textsc{Geary, R.~C.} (1954): \enquote{The contiguity ratio and statistical
  mapping,} \emph{The Incorporated Statistician}, 5, 115--146.

\bibitem[\protect\citeauthoryear{Gelb}{Gelb}{2021{\natexlab{a}}}]{spnet:R}
\textsc{Gelb, J.} (2021{\natexlab{a}}): \enquote{spNetwork, a package for
  network kernel density estimation,} \emph{The R Journal}.

\bibitem[\protect\citeauthoryear{Gelb}{Gelb}{2021{\natexlab{b}}}]{spnet:git}
---\hspace{-.1pt}---\hspace{-.1pt}--- (2021{\natexlab{b}}): \emph{spNetwork:
  Spatial Analysis on Network}.

\bibitem[\protect\citeauthoryear{Getis and Ord}{Getis and
  Ord}{1992}]{geatisord}
\textsc{Getis, A. and J.~K. Ord} (1992): \enquote{The Analysis of Spatial
  Association by Use of Distance Statistics,} \emph{Geographical Analysis}, 24,
  189--206.

\bibitem[\protect\citeauthoryear{Lee}{Lee}{2001}]{lee2001}
\textsc{Lee, S.-I.} (2001): \enquote{Developing a bivariate spatial association
  measure: An integration of Pearson's r and Moran's I,} \emph{Journal of
  Geographical Systems}, 3, 369--385.

\bibitem[\protect\citeauthoryear{Lu, Sun, Xu, Harris, and Charlton}{Lu
  et~al.}{2018}]{Shp2graph}
\textsc{Lu, B., H.~Sun, M.~Xu, P.~Harris, and M.~Charlton} (2018):
  \enquote{{Shp2graph}: Tools to Convert a Spatial Network into an Igraph Graph
  in R,} \emph{ISPRS International Journal of Geo-Information}, 7, 293.

\bibitem[\protect\citeauthoryear{Magzhan and Jani}{Magzhan and
  Jani}{2013}]{Magzhan:etal:2013}
\textsc{Magzhan, K. and H.~Jani} (2013): \enquote{A Review And Evaluations Of
  Shortest Path Algorithms,} \emph{International journal of scientific \&
  technology research}, 2.

\bibitem[\protect\citeauthoryear{Moradi, Cronie, and Mateu}{Moradi
  et~al.}{2020}]{stlnpp}
\textsc{Moradi, M., O.~Cronie, and J.~Mateu} (2020): \emph{stlnpp:
  Spatio-temporal analysis of point patterns on linear networks}.

\bibitem[\protect\citeauthoryear{Moran}{Moran}{1950}]{moran}
\textsc{Moran, P. A.~P.} (1950): \enquote{Notes on continuous stochastic
  phenomena,} \emph{Biometrika}, 37, 17--23.

\bibitem[\protect\citeauthoryear{Okabe and Yamada}{Okabe and
  Yamada}{2001}]{https://doi.org/10.1111/j.1538-4632.2001.tb00448.x}
\textsc{Okabe, A. and I.~Yamada} (2001): \enquote{The K-Function Method on a
  Network and Its Computational Implementation,} \emph{Geographical Analysis},
  33, 271--290.

\bibitem[\protect\citeauthoryear{Okabe, Yomono, and Kitamura}{Okabe
  et~al.}{1995}]{https://doi.org/10.1111/j.1538-4632.1995.tb00341.x}
\textsc{Okabe, A., H.~Yomono, and M.~Kitamura} (1995): \enquote{Statistical
  Analysis of the Distribution of Points on a Network,} \emph{Geographical
  Analysis}, 27, 152--175.

\bibitem[\protect\citeauthoryear{Ontario}{Ontario}{2006}]{ontario:2006}
\textsc{Ontario, P.} (2006): \enquote{Ontario Road Network,} Data retrieved
  from Ontario Ministry of Natural Resources: Peterborough, ON, Canada,
  \url{http://www.geographynetwork.ca/website/orn/viewer.htm}.

\bibitem[\protect\citeauthoryear{{QGIS Development Team}}{{QGIS Development
  Team}}{2022}]{QGIS:software}
\textsc{{QGIS Development Team}} (2022): \emph{QGIS Geographic Information
  System}, QGIS Association.

\bibitem[\protect\citeauthoryear{{R Core Team}}{{R Core Team}}{2021}]{RCore}
\textsc{{R Core Team}} (2021): \emph{R: A Language and Environment for
  Statistical Computing}, R Foundation for Statistical Computing, Vienna,
  Austria.

\bibitem[\protect\citeauthoryear{Rahman}{Rahman}{2017}]{Rahman2017}
\textsc{Rahman, M.~S.} (2017): \emph{Paths, Cycles, and Connectivity}, Cham:
  Springer International Publishing, 31--46.

\bibitem[\protect\citeauthoryear{Rasmussen and Christensen}{Rasmussen and
  Christensen}{2021}]{Rasmussen2021}
\textsc{Rasmussen, J.~G. and H.~S. Christensen} (2021): \enquote{Point
  Processes on Directed Linear Networks,} \emph{Methodology and Computing in
  Applied Probability}, 23, 647--667.

\bibitem[\protect\citeauthoryear{Redlands}{Redlands}{2011}]{ArcGis:software}
\textsc{Redlands, C. E. S. R.~I.} (2011): \emph{ArcGIS Geographic Information
  System}.

\bibitem[\protect\citeauthoryear{Schneble}{Schneble}{2021}]{geonet}
\textsc{Schneble, M.} (2021): \emph{geonet: Intensity Estimation on Geometric
  Networks with Penalized Splines}, r package version 0.6.0.

\bibitem[\protect\citeauthoryear{{van Lieshout}}{{van
  Lieshout}}{2018}]{LieshoutNets}
\textsc{{van Lieshout}, M. N.~M.} (2018): \enquote{Nearest-neighbour Markov
  point processes on graphs with Euclidean edges,} \emph{Advances in Applied
  Probability}, 50, 1275--1293.

\end{thebibliography}
\end{document}